\documentclass[11pt]{article}
\usepackage{cite}
\usepackage{amsmath,amsfonts,amssymb}
\usepackage[small,bf,hang]{caption}

\def\hybrid{
        \topmargin -20pt
        \oddsidemargin 0pt
        \headheight 0pt \headsep 0pt
        \textwidth 6.25in 
        \textheight 9.5in 
        \marginparwidth .875in
        \parskip 5pt plus 1pt \jot = 1.5ex}

\hybrid

 \csname
@addtoreset\endcsname{equation}{section}

\def\moth{\mathsurround=0pt}
\newdimen\zo \zo=0pt

\def\tick{\leaders\hrule height 0.5ex depth 0pt \hskip 0.5pt}
\def\upboxfill{$\moth \setbox\zo\hbox{\tick}%
  \hskip 3pt\hbox to 0pt{$\tick$\hss}\hrulefill \hbox to 7.5pt{$\tick$\hss}$}

\def\dtick{\leaders\hrule height .34pt depth 0.5ex \hskip 0.5pt}
\def\downboxfill{$\moth \setbox\zo\hbox{\dtick}%
  \hskip 2pt\hbox to 0pt{$\dtick$\hss}\hrulefill \hbox to 2pt{$\dtick$\hss}$}

\def\bec{\begin{center}}
\def\ec{\end{center}}

\def\be{\begin{equation}}
\def\ee{\end{equation}}
\def\bea{\begin{eqnarray}}
\def\eea{\end{eqnarray}}
\def\ba{\begin{array}}
\def\ea{\end{array}}

\begin{document}

\begin{titlepage}
\rightline{March 2011}
\rightline{\tt MIT-CTP-4223}
\begin{center}
\vskip 2.5cm
{\Large \bf {
Double Field Theory Formulation of Heterotic Strings
}}\\
\vskip 1.0cm
{\large {Olaf Hohm and Seung Ki Kwak}}
\vskip 1cm
{\it {Center for Theoretical Physics}}\\
{\it {Massachusetts Institute of Technology}}\\
{\it {Cambridge, MA 02139, USA}}\\
ohohm@mit.edu, sk$\_$kwak@mit.edu

\vskip 2.5cm
{\bf Abstract}
\end{center}

\vskip 0.1cm

\noindent
\begin{narrower}

We extend the recently constructed double field theory formulation of the low-energy theory
of the closed bosonic string to the heterotic string. The action can be written in terms of a generalized metric
that is a covariant tensor under $O(D,D+n)$, where $n$ denotes the number of
gauge vectors, and $n$ additional coordinates are introduced together with a covariant
constraint that locally removes these new coordinates. For the abelian subsector,
the action takes the same structural form as for the bosonic string, but based on the
enlarged generalized metric, thereby featuring a global $O(D,D+n)$ symmetry.
After turning on non-abelian gauge couplings, this global symmetry is broken, but
the action can still be written in a fully $O(D,D+n)$ covariant fashion, in
analogy to similar constructions in gauged supergravities.

\end{narrower}

\end{titlepage}

\newpage

\tableofcontents
\baselineskip=16pt

\section{Introduction and Overview}
Recently, a `double field theory' extension of the low-energy theory of closed bosonic
strings has been found, in which the T-duality group $O(D,D)$ is realized
as a global symmetry by virtue of doubling the coordinates
\cite{Hull:2009mi,Hull:2009zb,Hohm:2010jy,Hohm:2010pp} (see also
\cite{Kwak:2010ew,Hohm:2010xe,Berman:2010is,West:2010ev,Jeon:2010rw,Hohm:2011dz}
and \cite{Hohm:2011gs} for a review). More precisely,
the conventional low-energy effective action for the metric $g_{ij}$, the Kalb-Ramond
2-form $b_{ij}$ and the dilaton $\phi$,
 \bea\label{original}
  S \ = \ \int dx \sqrt{g}e^{-2\phi}\left[R+4(\partial\phi)^2-\frac{1}{12}H^2\right]\,,
 \eea
where $H_{ijk}=3\partial_{[i}b_{jk]}$, can be extended to an action written in terms of the
`generalized metric'
 \be\label{genmetric}
  {\cal H}^{MN} \ = \  \begin{pmatrix}    g_{ij}-b_{ik}g^{kl}b_{lj} & b_{ik}g^{kj}\\[0.5ex]
  -g^{ik}b_{kj} & g^{ij}\end{pmatrix}\;,
 \ee
and an $O(D,D)$ invariant dilaton $d$ defined by $e^{-2d}=\sqrt{g}e^{-2\phi}$.
Here, $M,N,\ldots=1,\ldots,2D$ are fundamental $O(D,D)$ indices, and the fields
have been grouped such that ${\cal H}^{MN}$ transforms covariantly under this group.
One can think of ${\cal H}$ as a (constrained) metric on the doubled space with
coordinates $X^{M}=(\tilde{x}_{i},x^{i})$, and all fields are assumed to depend
on the doubled coordinates.
The action then  takes a manifestly $O(D,D)$ invariant form and reads
 \bea\label{Hactionx}
 \begin{split}
  S \ = \ \int dx d\tilde{x}\,e^{-2d}~\Big(~&\frac{1}{8}\,{\cal H}^{MN}\partial_{M}{\cal H}^{KL}
  \,\partial_{N}{\cal H}_{KL}-\frac{1}{2}{\cal H}^{MN}\partial_{N}{\cal H}^{KL}\,\partial_{L}
  {\cal H}_{MK}\\
  &-2\,\partial_{M}d\,\partial_{N}{\cal H}^{MN}+4{\cal H}^{MN}\,\partial_{M}d\,
  \partial_{N}d~\Big)\,,
 \end{split}
 \eea
with derivatives $\partial_{M}=(\tilde{\partial}^{i},\partial_{i})$.
This action is also invariant under gauge transformations parametrized by $\xi^{M}=(\tilde{\xi}_{i},\xi^{i})$,
which take the form of `generalized Lie derivatives' $\widehat{\cal L}_{\xi}$,
 \be\label{manifestH}
 \begin{split}
  \delta_{\xi}{\cal H}^{MN} \ &= \  \widehat{\cal L}_{\xi} {\cal H}^{MN} \ \equiv  \ \xi^{P}\partial_{P}{\cal H}^{MN}
  +\big(\partial^{M}\xi_{P} -\partial_{P}\xi^{M}\big)\,{\cal H}^{PN}
  +
  \big(\partial^{N}\xi_{P} -\partial_{P}\xi^{N}\big)\,{\cal H}^{MP}\;, \\
  \delta d \ &= \ \xi^M \partial_M d - {1\over 2}  \partial_M \xi^M \,,
 \end{split}
 \ee
where indices are raised and lowered with the $O(D,D)$ invariant metric
 \be
  \eta_{MN} \ = \ \begin{pmatrix}
    0&1 \\1&0 \end{pmatrix}\;.
 \ee
We can think of the dilaton $d$ as a generalized density.
The gauge invariance and thus the consistency of the action (\ref{Hactionx}) requires the
following $O(D,D)$ covariant constraints
 \be\label{ODDconstr}
   \partial^{M}\partial_{M}A \ = \ \eta^{MN}\partial_{M}\partial_{N}A \ = \ 0\;, \qquad
   \partial^{M}A\,\partial_{M}B \ = \ 0\;,
 \ee
for arbitrary fields and parameters $A,B$. The first condition is the level-matching
condition for the massless fields in closed string theory.
The second condition is a stronger constraint that requires also all
possible products to be annihilated by $\partial^{M}\partial_{M}$.
This strong constraint implies that locally there is always an $O(D,D)$
transformation that rotates into a T-duality frame in which the fields depend only on
half of the coordinates, e.g., being independent of the $\tilde{x}_{i}$.

If the tilde coordinates $\tilde{x}_{i}$ are set to zero, the action (\ref{Hactionx}) reduces to
the low-energy action (\ref{original}), as required. Moreover, if these coordinates are set
to zero in (\ref{manifestH}), the gauge transformations reduce to the familiar diffeomorphisms
generated by $\xi^{i}$ and the Kalb-Ramond gauge transformations generated by $\tilde{\xi}_{i}$.

In this paper we are concerned with the extension of the above construction to the
heterotic string \cite{Gross:1985fr}.
In its low-energy limit, this theory is described by an effective two-derivative action whose  
bosonic terms extend (\ref{original})
by $n$ non-abelian
gauge fields $A_{i}{}^{\alpha}$, $\alpha=1,\ldots,n$, \cite{Gross:1985rr},
 \be\label{hetaction}
   S \ = \
   \int dx \sqrt{g} e^{-2 \phi}
   \bigg[ R + 4 (\partial\phi)^2
   - \frac{1}{12} \hat{H}^{ijk} \hat{H}_{ijk}
   - \frac{1}{4}F^{ij \alpha} F_{ij \alpha} \bigg] \, ,
 \ee
where
 \be
  F_{ij}{}^{\alpha} \ = \ \partial_i A_j{}^\alpha - \partial_j A_i{}^\alpha+g_0 \big[ A_{i},A_{j}\big]^{\alpha}\;
 \ee
is the non-abelian field strength of the gauge vectors, and the field strength of the $b$-field
gets modified by a Chern-Simons 3-form,
 \be\label{modH}
  \hat{H}_{ijk} \ = \ 3\left(
  \partial_{[i} b_{jk]} -
  \kappa_{\alpha\beta} A_{[i}{}^{\alpha}\Big(\partial_j A_{k]}{}^{\beta}
  +\tfrac{1}{3}g_0 \big[ A_{j},A_{k]}\big]^{\beta}\Big)\right)\;.
 \ee
Here $g_0$ denotes the gauge coupling constant and $\kappa_{\alpha\beta}$ is the invariant Cartan-Killing form.
With the gauge field transforming as
 \be
  \delta_{\Lambda}A_{i}{}^{\alpha} \ = \ \partial_{i}\Lambda^{\alpha}+g_0 \big[A_{i},\Lambda\big]^{\alpha}\;,
 \ee
the $b$-field transforms under $\Lambda^{\alpha}$ as
 \be\label{newdelb}
  \delta_{\Lambda}b_{ij} \ = \ \frac{1}{2}\big(\partial_{i}A_{j}{}^{\alpha}-\partial_{j}A_{i}{}^{\alpha}\big)
  \,\Lambda_{\alpha}\;,
 \ee
such that (\ref{modH}) is invariant.  At the level of the classical supergravity action, the gauge
group is arbitrary, but in heterotic string theory it is either $SO(32)$ or $E_8\times E_8$.

In sec.~2  we show that for the abelian subsector the double field theory
extension of the heterotic string is straightforward.
To this end, the coordinates are further extended by $n$ extra coordinates
$y^{\alpha}$ and, correspondingly, the generalized metric (\ref{genmetric}) is enlarged
to a $(2D+n)\times (2D+n)$ matrix that naturally
incorporates the additional fields $A_{i}{}^{\alpha}$ in precise analogy to the coset structure
appearing in dimensional reductions. This suggests an enhancement of the global symmetry
to $O(D,D+n)$. Indeed, if we formally keep the action (\ref{Hactionx})
and the form of the gauge transformations (\ref{manifestH}), but with respect to the
enlarged ${\cal H}^{MN}$, we
obtain precisely the (abelian subsector of the)
required action (\ref{hetaction}) and the correct gauge transformations
in the limit that the new coordinates are set to zero. In this construction, the number $n$ of
new coordinates is not constrained, but the case relevant for heterotic string theory is $n=16$,
where the $y^{\alpha}$ can be thought of as the coordinates of the internal torus corresponding
to the Cartan subalgebra of $SO(32)$ or $E_8\times E_8$.

In sec.~3 we turn to  the non-abelian extension. In this case the group $O(D,D+n)$ is broken.
More precisely, the reduction of the low-energy effective action (i.e., of heterotic supergravity)
on a torus $T^D$ gives rise to a theory with a global $O(D,D+n)$ symmetry only
in the abelian limit $g_0\rightarrow 0$ \cite{Maharana:1992my}.
Remarkably, however, we find that the action can be extended to incorporate the non-abelian
gauge couplings in a way that formally preserves $O(D,D+n)$, where $n$ equals the
dimension of the full gauge group. 
We write the extended action in terms of
a tensor $f^{M}{}_{NK}$, which encodes the structure constants of the gauge group, and
the generalized metric ${\cal H}^{MN}$. The consistency of
this construction requires a number of $O(D,D+n)$-covariant constraints on $f^{M}{}_{NK}$. Apart from
standard constraints like the Jacobi identities, there is one novel differential constraint
in addition to (\ref{ODDconstr}), which reads
 \be\label{fdelconstr0}
  f^{M}{}_{NK}\,\partial_{M} \ = \ 0 \;.
 \ee
Moreover, the gauge variations parametrized by $\xi^{M}$
get deformed by $f^{M}{}_{NK}$ in that, say, a `vector' $V^{M}$
transforms as
 \be\label{newgauge}
  \delta_{\xi}V^{M} \ = \ \widehat{\cal L}_{\xi}V^{M}-\xi^{K}f^{M}{}_{KL}V^{L}\;,
 \ee
where $ \widehat{\cal L}_{\xi}$ denotes the generalized Lie derivative as in (\ref{manifestH}).
Thus, the $\xi^{M}$ gauge transformations represent a curious mix between
diffeomorphism-like symmetries (which simultaneously treat each index as upper and lower index)
and the adjoint rotations with respect to some Lie group. The invariance of the action
under these deformed gauge transformations then requires new couplings to be added to
(\ref{Hactionx}), whose Lagrangian reads (without the $e^{-2d}$ prefactor)
 \be\label{fLagrang}
 \begin{split}
  {\cal L}_{f} \ = \  &-\frac{1}{2}f^{M}{}_{NK}\,{\cal H}^{NP}{\cal H}^{KQ}\partial_{P}{\cal H}_{QM}\\
  &-\frac{1}{12} f^{M}{}_{KP}f^{N}{}_{LQ}{\cal H}_{MN}{\cal H}^{KL}{\cal H}^{PQ}
  -\frac{1}{4}f^{M}{}_{NK}f^{N}{}_{ML}{\cal H}^{KL}-\frac{1}{6}f^{MNK}f_{MNK}\;.
 \end{split}
 \ee

Despite the $O(D,D+n)$ covariant form of the action, any
non-vanishing choice for the $f^{M}{}_{NK}$ will actually break the symmetry
to the subgroup that leaves this tensor invariant, because $f^{M}{}_{NK}$ is not a dynamical field
and therefore does not transform under the T-duality group.
For instance, if we choose $f^{M}{}_{NK}$
to be non-vanishing only for the components $f^{\alpha}{}_{\beta\gamma}$ that are
the structure constants of a semi-simple Lie group $G$, the remaining
symmetry will be $O(D,D)\times G$, where $G$ is the rigid subgroup of the gauge group.
In this case, the new couplings (\ref{fLagrang}) precisely constitute the non-abelian
gauge couplings required by (\ref{hetaction}), while the gauge variations (\ref{newgauge})
evaluated for ${\cal H}^{MN}$ reduce to the non-abelian Yang-Mills transformations.

It should be stressed that the abelian and non-abelian cases are conceptually quite different.
The abelian case is closely related to the original construction in
\cite{Hull:2009mi}. Specifically, if we choose $n=16$, the constraint (\ref{ODDconstr}) can be interpreted as
a stronger form of the level-matching condition. Moreover, the winding coordinates $\tilde{x}_{i}$ and
the $y^{\alpha}$ have a direct interpretation in the full string theory.
In contrast, the non-abelian case requires the new constraint (\ref{fdelconstr0}),
which has no obvious interpretation in string theory, and
formally we introduce as many new
coordinates as the dimension of the gauge group, i.e., $n=496$ for the case relevant to
heterotic string theory. However, the number $n$ is a free parameter at the level of
the double field theory constructions discussed here,
and therefore we will not introduce different notations for $n$ in the two cases.

We note that the constraint (\ref{fdelconstr0}) effectively
removes the dependence on (some of) the extra coordinates. More precisely,
a subtle interplay between the constraints (\ref{ODDconstr}) and (\ref{fdelconstr0})
and the unbroken part of the T-duality group guarantees locally independence
on the `unphysical' coordinates, as we will discuss in sec.~4.
It is amusing to note that this construction has a superficial similarity to attempts
in the early literature on heterotic string theory that aimed at realizing this theory through some
Kaluza-Klein type reduction from $496+10$ dimensions \cite{Duff:1985cm,Duff:1986ya},
but the details, in particular the physical
interpretation of the extra coordinates, appear to be different.
(See also the more recent work \cite{Andriot:2011iw}, which has some relevance for
the abelian case discussed in sec.~2.)

Interestingly, the results on the non-abelian case
are analogous to constructions of gauged supergravities based on the
so-called embedding tensor formalism (see \cite{Samtleben:2008pe} for a review and
references therein). In this formalism, the deformation of an
ungauged supergravity with a certain duality group $G$ into a gauged supergravity
is parametrized by the embedding tensor that is formally a tensor under $G$ and which is
the analogue of the tensor $f^{M}{}_{NK}$ above.
Even though the $G$-invariance is ultimately broken for any choice of (non-vanishing)
embedding tensor,
all couplings induced by the gauging can be written in a $G$-covariant fashion.
In particular, the scalar potential takes a form that is precisely analogous to the
terms in the second line of (\ref{fLagrang}). In gauged supergravity, however, the
exact form of these couplings
can only be determined by supersymmetry. It is remarkable, therefore,  that in the
construction to be discussed in this paper, the couplings (\ref{fLagrang}) are uniquely
determined by the bosonic symmetries (\ref{newgauge}) (apart from the last term
which is constant and thus separately gauge invariant).

The original construction of double field theory is closely related to a frame-like
geometrical formalism developed by Siegel in important independent work
\cite{Siegel:1993th,Siegel:1993xq}.
The precise relation to the formulation in terms of a generalized metric
is by now well-understood both at the level of the symmetry transformations \cite{Hohm:2010pp}
and the action \cite{Hohm:2010xe}. Siegel's formalism as presented in \cite{Siegel:1993th}
is already adapted to include the abelian subsector of the heterotic theory.
Using the recent
results of \cite{Hohm:2010xe}, it is straightforward to verify the equivalence of
this formalism with the generalized metric formulation in the abelian limit, which we
do in sec.~5. Moreover, the formulation of \cite{Siegel:1993th} also 
allows for supersymmetric extensions. We therefore expect a supersymmetric 
version of the formulation discussed here to be possible. This we will leave, however, for future work,
and we stress that whenever we refer in this paper to the heterotic string we mean, more precisely,
the bosonic sector of the low-energy action. 
Finally, in the conclusions to the proceedings of \textit{Strings\hspace{0.1em}'93}
\cite{Siegel:1993bj},
Siegel also mentions the extension to the non-abelian case, with a deformation of
the gauge variations as in (\ref{newgauge}) and a corresponding adaptation
of the frame formalism, which we will discuss in detail in sec.~5.

\section{Double field theory with abelian gauge fields}
In this section we introduce the double field theory formulation for the abelian subsector of
the low-energy theory of the heterotic string. We first define the enlarged generalized metric
and then show that the action (\ref{Hactionx}) and the gauge transformations (\ref{manifestH})
reduce to the required form when the dependence on the new coordinates is dropped.

\subsection{Conventions and generalized metric}
The coordinates are grouped according to
 \be
  X^M  \ = \ \left(\,\tilde{x}_i ,\, x^i ,\, y^\alpha\,\right)\;,
 \ee
which transforms as a fundamental $O(D,D+n)$ vector,
 \be
  X^{\prime M} \ = \ h^{M}{}_{N} \,X^{N}\;, \qquad
  h \ \in \ O(D,D+n)\;.
 \ee
Here, $O(D,D+n)$ is the group leaving the metric of
signature $(D,D+n)$ invariant,
 \be
  \eta^{MN} \ = \ h^{M}{}_{P}\,h^{N}{}_{Q}\,\eta^{PQ}\;,
 \ee
where
\be\label{neweta}
\eta_{MN}  \ = \  \begin{pmatrix} \eta^{ij} &  \eta^{i}{}_{j} & \eta^{i}{}_{\beta} \\ \eta_{i}{}^j & \eta_{ij} & \eta_{i \beta} \\ \eta_{\alpha}{}^{ j}  & \eta_{\alpha j} & \eta_{\alpha \beta} \end{pmatrix} \ = \ \begin{pmatrix} 0 & 1 & 0 \\ 1 & 0 &0 \\ 0 & 0 & \kappa \end{pmatrix}\;.
\ee
Here, we introduced $\kappa$ to denote the matrix corresponding to the Cartan-Killing
metric of the gauge group. In the present abelian case, this is simply given by the unit matrix,
$\kappa_{\alpha\beta}=\delta_{\alpha\beta}$, but we kept the notation more general for the later
extension to the non-abelian case.

According to these index conventions, the derivatives and gauge parameters are
\begin{eqnarray}\label{newparameter}
\partial_M &=& \big(\, \tilde{\partial}^i , \partial_i , \partial_\alpha\,\big)\;, \qquad
\xi^M \ = \ \big(\,\tilde{\xi}_i, \xi^i , \Lambda^\alpha\,\big) \, ,
\end{eqnarray}
which combines the gauge parameters of diffeomorphism, Kalb-Ramond and abelian
gauge transformations into an $O(D,D+n)$ vector.
The strong constraint (\ref{ODDconstr}) reads explicitly
\begin{eqnarray}\label{strong1}
\partial_M \partial^M A &=&  2  \tilde{\partial}^i \partial_i A + \partial_{\alpha} \partial^{\alpha} A \ = \  0 \ ,
\\\label{strong2}
\partial_M A\, \partial^M B & =& \tilde{\partial}^i A\,  \partial_i B + \partial_i A \,\tilde{\partial}^i B
+ \partial_\alpha A \,\partial^\alpha B \ = \ 0 \,\ ,
\end{eqnarray}
for arbitrary fields and gauge parameters $A$ and $B$.
As for the bosonic theory, this constraint is a stronger version of the level-matching condition
and it implies that locally there is always
an $O(D,D+n)$ transformation that rotates into a frame in which the fields depend only
on the $x^{i}$. We discuss this in more detail in sec.~4.

Next, we introduce the extended form of the generalized metric ${\cal H}^{MN}$
and require that it transforms covariantly under $O(D,D+n)$ ,
 \be\label{genBuscher}
  {\cal H}^{\, \prime MN}(X^{\prime}) \ = \ h^{M}{}_{P}\,h^{N}{}_{Q}\,{\cal H}^{PQ}(X)\;, \qquad
  d^{\prime}(X^{\prime}) \ = \ d(X)\;.
 \ee
In analogy to the structure encountered in dimensionally reduced theories
\cite{Maharana:1992my}, we make the ansatz
\be\label{newH}
{\cal H}_{MN}  =    \begin{pmatrix}   {\cal H}^{ij} &  {\cal H}^{i}{}_{j} & {\cal H}^{i}{}_{ \beta} \\ {\cal H}_{i}{}^j & {\cal H}_{ij} & {\cal H}_{i \beta} \\ {\cal H}_{\alpha}{}^{j}  & {\cal H}_{\alpha j} & {\cal H}_{\alpha \beta}
\end{pmatrix}
  =  \begin{pmatrix} g^{ij} & - g^{ik} c_{kj}  & - g^{ik} A_{k \beta} \\
- g^{jk} c_{ki}  & g_{ij} + c_{ki} g^{kl} c_{lj} + A_{i}{}^{\gamma} A_{j \gamma} & c_{ki} g^{kl} A_{l \beta}
 + A_{i \beta}  \\
 - g^{jk} A_{k \alpha} & c_{kj} g^{kl} A_{l \alpha} + A_{ j \alpha} & \kappa_{\alpha \beta}
 + A_{k \alpha} g^{kl} A_{l \beta}
\end{pmatrix} \ ,
\ee
where
gauge group indices $\alpha,\beta,\ldots$ are raised and lowered with $\kappa_{\alpha\beta}$, and
 \be\label{c}
  c_{ij} \ = \ b_{ij} + \frac{1}{2} A_{i}{}^{\alpha} A_{j}{}_{\alpha} \;.
 \ee
The generalized metric defined like this is still symmetric,
${\cal H}_{MN} = {\cal H}_{NM}$. Raising all indices with $\eta^{MN}$, we obtain
\be\label{Hgenup}
{\cal H}^{MN} =    \begin{pmatrix}      {\cal H}_{ij} & {\cal H}_{i}{}^j & {\cal H}_{i}{}^{\beta} \\     {\cal H}^{i}{}_{j} & {\cal H}^{ij} & {\cal H}^{i \beta} \\   {\cal H}^{\alpha}{}_j & {\cal H}^{\alpha j} & {\cal H}^{\alpha \beta}
\end{pmatrix}
  = \begin{pmatrix}   g_{ij} + c_{ki} g^{kl} c_{lj} + A_{i}{}^{\gamma} A_{j \gamma} & - g^{jk} c_{ki} & c_{ki} g^{kl} A_{l}{}^{\beta} + A_{i}{}^{\beta}  \\
  - g^{ik} c_{kj} & g^{ij}   & - g^{ik} A_{k}{}^{\beta} \\
  c_{kj} g^{kl} A_{l}{}^{\alpha} + A_{j}{}^{\alpha} & - g^{jk} A_{k}{}^{\alpha} & \kappa^{\alpha \beta}
  + A_{k}{}^{\alpha} g^{kl} A_{l}{}^{\beta}
\end{pmatrix} \ .
\ee
This is the inverse of (\ref{newH}), and so the generalized metric satisfies the constraint
${\cal H}^{MK} {\cal H}_{KN} = \delta^{M}{}_N$. This implies that, viewed as a matrix, it
is an element of $O(D,D+n)$ in that it satisfies
\be
{\cal H}^{-1} \ = \ \eta\, {\cal H}\, \eta \, .
\ee
The $O(D,D+n)$ action (\ref{genBuscher}) defines the generalized Buscher rules for the abelian subsector
of heterotic string theory.

\subsection{Gauge symmetries}
We turn now to the gauge transformations of the component fields that follow from the
extended form of the generalized metric (\ref{Hgenup}) and the generalized Lie derivatives
(\ref{manifestH}) with respect to the extended parameter (\ref{newparameter}).
Specifically, we verify that for $\tilde{\partial}^i = \partial_{\alpha} = 0$ the gauge transformations
of the component fields take the required form.

For the gauge variation of ${\cal H}^{ij}$ we find
\begin{eqnarray}
\delta_{\xi} {\cal H}^{ij} &=& \delta_{\xi} g^{ij}  \ = \   \  \xi^k \partial_k {\cal H}^{ij}  - \partial^P \xi^i \, {\cal H}_{P}{}^{j}   - \partial^P \xi^j \, {\cal H}^{i}{}_{P}
\\ \nonumber &=& \xi^k \partial_k g^{ij}  - \partial_k \xi^i \, g^{kj}   - \partial_k \xi^j \, g^{ik} \ = \ {\cal L}_{\xi} g^{ij} \, ,
\end{eqnarray}
i.e., the metric $g_{ij}$ transforms as expected with the Lie derivative under
diffeomorphisms parametrized by $\xi^i$ and is inert under the other gauge symmetries.
For the component ${\cal H}^{i \beta }$ we infer
\begin{eqnarray}
\delta_{\xi} {\cal H}^{i \beta } &=& \delta_{\xi} \big( - g^{ik} A_{k}{}^{\beta}\big) \ = \  \xi^k \partial_k {\cal H}^{i \beta}  - \partial^P \xi^i \, {\cal H}_{P}{}^{\beta}   - \partial^P \xi^{\beta} \, {\cal H}^{i}{}_{P}
\\ \nonumber &=& \xi^k \partial_k {\cal H}^{i \beta}  - \partial_k \xi^i \, {\cal H}^{k \beta}   - \partial_k \xi^{\beta} \, {\cal H}^{ik}
\\ \nonumber &=& \xi^k \partial_k  \big( - g^{il} A_{l}{}^{\beta}\big) - \partial_k \xi^i \,
\big( - g^{kl} A_{l}{}^{\beta}\big) - \partial_k \Lambda^{\beta} g^{ik}
\\ \nonumber &=& {\cal L}_{\xi} \big( - g^{ik} A_{k}{}^{\beta}\big) -  g^{ik} \partial_k \Lambda^{\beta} \;.
\end{eqnarray}
Together with the form of $\delta_{\xi}g^{ij}$ determined above, this implies for the gauge vectors
 \be
   \delta_{\xi} A_{k}{}^{\beta} \ = \ {\cal L}_{\xi} A_{k}{}^{\beta} + \partial_k \Lambda^{\beta}\;,
 \ee
which
represents the expected diffeomorphism and abelian gauge transformation.
Finally, for the component ${\cal H}^{i}{}_{j}$ we derive
\begin{eqnarray}
\delta_{\xi}  {\cal H}^{i}{}_{j}   &=& \delta_{\xi} \big( - g^{ik} c_{kj}\big) \ = \ \xi^k \partial_k  {\cal H}^{i}{}_{j} - \partial^P \xi^i \, {\cal H}_{P j} + \big(\partial_j \xi^P - \partial^P \xi_j \big)  {\cal H}^{i}{}_{ P}
\\ \nonumber &=& \xi^k \partial_k {\cal H}^{i}{}_{j} -  \partial_k \xi^i \, {\cal H}^{k}{}_{j} + \partial_j \xi^k \, {\cal H}^{i}{}_{k} + \partial_j \tilde{\xi}_k \, {\cal H}^{ik} +  \partial_j \xi^{\beta} \, {\cal H}^{i}{}_{\beta} - \partial_k \tilde{\xi}_j \, {\cal H}^{ik}
\\ \nonumber &=& {\cal L}_{\xi} {\cal H}^{i}{}_{j} + \big( \partial_j  \tilde{\xi}_k - \partial_k \tilde{\xi}_j \big)
{\cal H}^{ik} +  \partial_j \xi^{\beta} \, {\cal H}^{i}{}_{\beta}
\\ \nonumber &=& {\cal L}_{\xi} \big( - g^{ik} c_{kj}\big) + \big( \partial_j  \tilde{\xi}_k - \partial_k \tilde{\xi}_j \big)
g^{ik} +  \partial_j \Lambda^{\beta} \,  \big( - g^{ik} A_{k \beta}\big) \, .
\end{eqnarray}
Using again the known form of the gauge transformation $\delta_{\xi}g^{ij}$, this implies
for the tensor defined in (\ref{c})
 \be
  \delta_{\xi} c_{ij} \ = \ {\cal L}_{\xi} c_{ij} + \big(\partial_i \tilde{\xi}_j - \partial_j  \tilde{\xi}_i \big)
  + A_{i \beta} \partial_j \Lambda^{\beta}\;.
 \ee
In order to derive the gauge transformation of $b_{ij}$,
we project this onto the symmetric and antisymmetric part,
\begin{eqnarray}
\delta_{\xi} c_{(ij)} &=& \delta_{\xi} \big(\tfrac{1}{2} A_{i \beta} A_{j}{}^{\beta}\big)
\ = \ {\cal L}_{\xi} \big(\tfrac{1}{2} A_{i \beta} A_{j}{}^{\beta}\big)
+ \tfrac{1}{2} \big( A_{i \beta} \partial_{j} \Lambda^{\beta} + A_{j \beta} \partial_{i} \Lambda^{\beta}\big) \, ,
\\ \delta_{\xi} c_{[ij]} &=& \delta_{\xi} b_{ij} \ = \ {\cal L}_{\xi} b_{ij}
+ \big(\partial_i \tilde{\xi}_j - \partial_j  \tilde{\xi}_i \big) + \tfrac{1}{2} \big( A_{i \beta} \partial_{j} \Lambda^{\beta} - A_{j \beta} \partial_{i} \Lambda^{\beta}  \big) \, .
\end{eqnarray}
The first equation is consistent with the gauge transformation of the gauge field as obtained above, while
the second equation yields the gauge transformation of $b_{ij}$.

To summarize, the gauge transformations in the limit
$\tilde{\partial}^i = \partial_{\alpha} = 0$ read
\begin{eqnarray}\label{finalgauge}
\delta g_{ij} &=& {\cal L}_{\xi} g_{ij}  \, , \\ \label{finalgauge2}
\delta A_{i}{}^{\alpha} &=& {\cal L}_{\xi} A_{i}{}^{\alpha} + \partial_i \Lambda^{\alpha} \, ,
\\ \label{finalgauge3}
\delta b_{ij} &=&  {\cal L}_{\xi} b_{ij}
+ \big(\partial_i \tilde{\xi}_j - \partial_j  \tilde{\xi}_i \big)
+ \frac{1}{2} \big( A_{i \alpha} \partial_{j} \Lambda^{\alpha}
- A_{j \alpha} \partial_{i} \Lambda^{\alpha}  \big) \, .
\end{eqnarray}
For metric and gauge vector, these give the expected result, but for $b_{ij}$ a
parameter redefinition is required in order to obtain (\ref{newdelb}). If we redefine
the one-form parameter $\tilde{\xi}_{i}$ according to
 \be\label{parare}
  \tilde{\xi}_{i}^{\prime} \ := \ \tilde{\xi}_{i} -\frac{1}{2}A_{i}{}^{\alpha} \Lambda_{\alpha}\;,
 \ee
the gauge variation of $b_{ij}$ becomes
 \be\label{gaugeinvvar}
   \delta b_{ij} \ = \ \partial_{i}\tilde{\xi}_{j}^{\prime}-\partial_{j}\tilde{\xi}_{i}^{\prime}
   +\frac{1}{2}F_{ij}{}^{\alpha}\Lambda_{\alpha}\;,
 \ee
with the abelian field strength $F_{ij}{}^{\alpha}$,
in accordance with (\ref{newdelb}).

We close this section with a brief discussion of the closure of the gauge transformations.
Using the form (\ref{manifestH}), one may verify that their commutator is given by
 \be
   \bigl[ \, \delta_{\xi_1}\,,  \delta_{\xi_2} \, \bigr]\,
   \ = \ - \delta_{[\xi_1, \xi_2]_{{}_{\rm C}}}   \,,
 \ee
where
\be
 \label{cbracketdef}
  \bigl[ \xi_1,\xi_2\bigr]_{\rm{C}}^{M} \ \equiv
   \ \xi_{1}^{N}\partial_{N}\xi_{2}^M -\frac{1}{2}\,  \xi_{1}^P\partial^{M}\xi_{2\,P}
   -(1\leftrightarrow 2)\;.
 \ee
This has been proved in \cite{Hohm:2010pp} in the original double field theory based on
the generalized metric (\ref{genmetric}), but since this derivation requires only the general form
of the gauge transformations (\ref{manifestH}) and the constraints (\ref{ODDconstr}),
this result immediately generalizes to the present case. In the original case,
this bracket (`C-bracket') reduces to the Courant bracket of generalized geometry for
$\tilde{\partial}=0$ \cite{Hull:2009zb,Tcourant,Hitchin,Gualtieri}.
Let us see how this generalizes after adding the $n$ additional components
for $\xi^{M}$. Setting now also $\partial_{\alpha}=0$, we obtain for the
various components of (\ref{cbracketdef})
 \begin{equation}\label{cbr1}
  \big(\bigl[ \xi_1,\xi_2\bigl]_{\rm C}\big)^{i} \ = \ \xi_1^{j}\partial_{j}\xi_2^i
  -\xi_2^{j}\partial_{j}\xi_1^i \ \equiv \ \bigl[\xi_1,\xi_2\bigl]^{i}\;,
 \end{equation}
which is unmodified and given by the usual Lie bracket,
 \begin{equation}\label{cbr2}
 \begin{split}
  \big(\bigl[ \xi_1,\xi_2\bigl]_{\rm C}\big)_{i}
  \ &= \ {\cal L}_{\xi_1}\tilde{\xi}_{2i}-{\cal L}_{\xi_2} \tilde{\xi}_{1i} -\frac{1}{2}\partial_{i}
  \big(\tilde{\xi}_{2j}\xi_1^j\big)+\frac{1}{2}\partial_{i}
  \big(\tilde{\xi}_{1j}\xi_2^j\big) -\frac{1}{2}\big(\Lambda_{1\alpha}\partial_{i}\Lambda_2{}^{\alpha}
  -\Lambda_{2\alpha}\partial_i\Lambda_1{}^{\alpha}\big)\;,
 \end{split}
 \end{equation}
which receives a new contribution involving $\Lambda$, and finally
 \be\label{cbr3}
  \big(\bigl[ \xi_1,\xi_2\bigl]_{\rm C}\big)^{\alpha} \ = \ \xi_1^{j}\partial_j \Lambda_2{}^{\alpha}-
  \xi_2^{j}\partial_j \Lambda_1{}^{\alpha}\;,
 \ee
which is the (antisymmetrized) Lie derivative of $\Lambda$. The Courant bracket  is defined
as a structure
on the direct sum of tangent and cotangent bundle over the space-time base manifold $M$,
$(T\oplus T^{*})M$, whose sections are formal sums $\xi+\tilde{\xi}$ of vectors and one-forms.
Thus, for the given generalization it is natural to consider a bundle that is further
extended to $T\oplus T^*\oplus V$, where we identify the sections of $V$ with the $\Lambda^{\alpha}$.
The sections of the total bundle are then written as $\xi+\tilde{\xi}+\Lambda$, and in this language, the
results (\ref{cbr1}), (\ref{cbr2}) and (\ref{cbr3}) can be summarized by
 \be\label{newbracket}
  \begin{split}
   \bigl[ \,\xi_1+\tilde{\xi}_1+\Lambda_1 \,,&\;  \xi_2+\tilde{\xi_2}+\Lambda_2\,\bigl] \ = \
   \bigl[ \xi_1,\xi_2\bigl] \\
   &+ {\cal L}_{\xi_1}\tilde{\xi}_2 - {\cal L}_{\xi_2}\tilde{\xi}_1
  -\frac{1}{2}d\big(i_{{\xi}_1}\tilde{\xi}_2-i_{{\xi}_2}\tilde{\xi}_1\big)
  -\frac{1}{2}\big(\langle\Lambda_1,d\Lambda_2\rangle-\langle\Lambda_2,d\Lambda_1\rangle\big)\\
  &+{\cal L}_{\xi_1}\Lambda_2-{\cal L}_{\xi_2}\Lambda_1\;,
 \end{split}
\ee
where
$\langle \Lambda_1,\Lambda_2\rangle=\kappa_{\alpha\beta}\Lambda_1^{\alpha}\Lambda_2^{\beta}$
denotes the inner product, and $i$ is the canonical product between vectors and one-forms.
Here, the term on the right-hand side in the first line represents the vector part, the terms in
the second line represent the one-form part, and finally the terms in the last line
represent the $V$-valued part. For $\Lambda=0$ this
reduces to the Courant bracket.

The bracket (\ref{newbracket}) implies in particular that the abelian gauge transformations
parametrized by $\Lambda^{\alpha}$ close into the gauge transformations of the 2-form.
This can also be confirmed directly from (\ref{finalgauge2}) and (\ref{finalgauge3}),
 \be\label{norbracket}
  \bigl[ \delta_{\Lambda_1},\delta_{\Lambda_2}\bigl]b_{ij} \ = \
  \delta_{\tilde{\xi}}b_{ij}\;, \qquad
  \tilde{\xi}_{i} \ = \ \frac{1}{2}\big(\Lambda_{1 \alpha}\partial_{i}\Lambda_2{}^{\alpha}
  -\Lambda_{2 \alpha}\partial_{i}\Lambda_1{}^{\alpha}\big)\;.
 \ee
We stress, however, that this result depends on a choice of basis for the gauge parameters.
In fact, after the parameter redefinition (\ref{parare}), the 2-form varies
into the gauge invariant field strength according to (\ref{gaugeinvvar}) and
thus the commutator trivializes.

\subsection{The action}
Let us now turn to the action (\ref{Hactionx}) applied to the extended form (\ref{Hgenup}) of
the generalized metric. We show that for
$\tilde{\partial}^i = \partial_{\alpha} = 0$ it reduces to the (abelian) low-energy action
(\ref{hetaction}) of the heterotic string.

The relevant terms in the action, setting
$\tilde{\partial}^i = \partial_{\alpha} = 0$, are given by
\begin{eqnarray}\label{redaction}
 \begin{split}
  S \ = \ \int dx  \,e^{-2d}~\Big(~&\frac{1}{8}\,{\cal H}^{ij}\partial_{i}{\cal H}^{KL}
  \,\partial_{j}{\cal H}_{KL}-\frac{1}{2}{\cal H}^{M i}\partial_{i}{\cal H}^{K j }\,\partial_{j}
  {\cal H}_{MK}\\
  &
  -2\,\partial_{i}d\,\partial_{j}{\cal H}^{i j}+4{\cal H}^{i j}\,\partial_{i}d\,
  \partial_{j}d~\Big)\,.
 \end{split}
\end{eqnarray}
The last two terms are unchanged as compared to the original case without gauge vectors
since the component ${\cal H}^{ij}=g^{ij}$ is unmodified.
Thus, we only need to examine the first two terms.
The first term reads
\begin{eqnarray}\nonumber
\frac{1}{8}{\cal H}^{ij}\partial_{i}{\cal H}^{KL}  \,\partial_{j}{\cal H}_{KL} &=&\frac{1}{4} \partial_i {\cal H}^{kl} \, \partial^i {\cal H}_{kl} + \frac{1}{4} \partial_i {\cal H}_{k}{}^{l} \, \partial^i {\cal H}^{k}{}_{l} + \frac{1}{2} \partial_i {\cal H}^{\alpha l} \, \partial^i {\cal H}_{\alpha l} + \frac{1}{8}\partial_i {\cal H}^{\alpha \beta} \, \partial^i {\cal H}_{\alpha \beta}
\\  &=& \frac{1}{4} \partial_i g^{lp} \, \partial^i \big(g_{lp} + c_{kp} g^{kq} c_{ql} + A_{l}{}^{\alpha} A_{p \alpha}\big) + \frac{1}{4} \partial_i \big(g^{lp} c_{pk}\big) \, \partial^i \big( g^{kq} c_{ql}\big)
\\ \nonumber &&   - \frac{1}{2} \partial_i \big( g^{lp} A_{p}{}^{\alpha}\big) \, \partial^i \big( c_{ql} g^{qk} A_{k \alpha} + A_{l \alpha} \big) + \frac{1}{8}\partial_i \big(A_{p}{}^\alpha g^{pl} A_{l}{}^\beta\big) \, \partial^i \big(A_{k \alpha} g^{kq} A_{q \beta}\big) \, .
\end{eqnarray}
After some work, this can be simplified to
\be\label{step01}
\frac{1}{8}\,{\cal H}^{ij}\,\partial_{i}{\cal H}^{KL}
  \,\partial_{j}{\cal H}_{KL} \ = \ \frac{1}{4} g^{ij}\,\partial_i g^{kl} \partial_j g_{kl}
  - \frac{1}{2} g^{ij} g^{kl} \partial_i A_{k \alpha} \, \partial_j A_{l}{}^{\alpha} - \frac{1}{4} \tilde{H}_{ijk} \tilde{H}^{ijk} \, ,
\ee
where $\tilde{H}_{ijk}=\partial_{i}b_{jk}-\partial_{i}A_{[j}{}^{\alpha}\,A_{k] \alpha}$.

Next we consider the second term in (\ref{redaction}), which yields
\begin{eqnarray}\nonumber
-\frac{1}{2}{\cal H}^{M i}\partial_{i}{\cal H}^{K j }\,\partial_{j} {\cal H}_{MK}  &=&
-\frac{1}{2}{\cal H}^{mi} \big( \partial_i {\cal H}^{kj} \, \partial_j {\cal H}_{mk} + \partial_i {\cal H}_{k}{}^{j} \, \partial_j {\cal H}_{m}{}^{k} + \partial_i {\cal H}^{\alpha j} \, \partial_j {\cal H}_{m \alpha} \big) \\ \nonumber
&&-\frac{1}{2}
 {\cal H}_{m}{}^{i} \big( \partial_i {\cal H}^{kj} \, \partial_j {\cal H}^{m}{}_{k} + \partial_i {\cal H}_{k}{}^{j} \, \partial_j {\cal H}^{mk}
+ \partial_i {\cal H}^{\alpha j} \, \partial_j {\cal H}^{m}{}_{\alpha} \big)\\ 
 &&-\frac{1}{2}
 {\cal H}^{\beta i} \big( \partial_i {\cal H}^{kj} \, \partial_j {\cal H}_{\beta k} + \partial_i {\cal H}_{k}{}^{j} \, \partial_j {\cal H}_{\beta}{}^{ k}  + \partial_i {\cal H}^{\alpha j} \, \partial_j {\cal H}_{\beta \alpha} \big) \, .
\end{eqnarray}
To simplify the evaluation of these terms, it is convenient to work out the following structures
separately,
\begin{eqnarray}
-\frac{1}{2}{\cal H}^{M i}\partial_{i}{\cal H}^{K j }\,\partial_{j} {\cal H}_{MK}  \big|_{(\partial g)^2} &=&
-\frac{1}{2}g^{ij} \partial_j g^{kl} \, \partial_l g_{ik} \, , \\
-\frac{1}{2}{\cal H}^{M i}\partial_{i}{\cal H}^{K j }\,\partial_{j} {\cal H}_{MK}  \big|_{(\partial g)^1} &=& 0 \, , \\
-\frac{1}{2}{\cal H}^{M i}\partial_{i}{\cal H}^{K j }\,\partial_{j} {\cal H}_{MK}  \big|_{(\partial g)^0} &=&
\frac{1}{2}g^{ik} g^{jl} \partial_i A_{l}{}^\alpha \, \partial_j A_{k \alpha} -\frac{1}{2} \tilde{H}_{ijk} \tilde{H}^{jki} \, .
\end{eqnarray}
Combining these three structures, we obtain
\be\label{step02}
-\frac{1}{2} {\cal H}^{M i}\partial_{i}{\cal H}^{K j }\,\partial_{j} {\cal H}_{MK}
\ = \ -\frac{1}{2} g^{ij} \partial_j g^{kl} \, \partial_l g_{ik} + \frac{1}{2} g^{ik} g^{jl} \partial_i A_{l}{}^\alpha \, \partial_j A_{k \alpha} - \frac{1}{4} \tilde{H}_{ijk} ( \tilde{H}^{jki} +\tilde{H}^{kij}) \, .
\ee
Finally, using (\ref{step01}) and (\ref{step02}), the reduced action (\ref{redaction})
can be written as
\be
\begin{split}
  S \ = \ \int dx  \,e^{-2d}~\Big(\,&  \frac{1}{4} g^{ij} \partial_i g^{kl} \partial_j g_{kl} -\frac{1}{2} g^{ij} \partial_j g^{kl} \, \partial_l g_{ik} -2\,\partial_{i}d\,\partial_{j}g^{i j}+4g^{ij}\,\partial_{i}d\,  \partial_{j}d \\
  &- \frac{1}{12} \hat{H}^2 - \frac{1}{4} F_{ij  \alpha} F^{ij  \alpha} \Big) \, .
\end{split}
\ee
Up to boundary terms, the terms in the first line are equivalent to the Einstein-Hilbert term coupled to the dilaton, compare eq.~(3.18) in \cite{Hohm:2010jy}. Thus, the reduced action
coincides precisely with (\ref{hetaction}).

\section{Non-abelian generalization}
In this section we generalize the previous results to non-abelian gauge groups.
This will be achieved by introducing a `duality-covariant'
form of the structure constants of the gauge group. While this object is not an invariant tensor
under $O(D,D+n)$ and so the T-duality group is no longer a proper symmetry, remarkably
the action and gauge transformations can still be written in an $O(D,D+n)$ invariant fashion.

\subsection{Duality-covariant structure constants}
We encode the structure constant in an object $f^{M}{}_{NK}$ that formally can be regarded
as a tensor under $O(D,D+n)$, even though it is ultimately fixed to be constant and thus not
to transform according to its index structure. To be specific, let us fix an $n$-dimensional
semi-simple Lie group $G$ whose Lie algebra has the
structure constants $f^{\alpha}{}_{\beta\gamma}$. Then we can define
\be\label{fform}
   f^{M}{}_{NK} \ = \  \left\{
  \begin{array}{l l}
    f^{\alpha}{}_{\beta\gamma} & \quad \text{if\; $(M,N,K)=(\alpha,\beta,\gamma)$}\\
    0 & \quad \text{else}\\
  \end{array} \right. \;.
\ee
This is \textit{not} an invariant tensor under $O(D,D+n)$, rather it will break this
symmetry to $O(D,D)\times G$. The advantage of this formulation is, however, that
the explicit form of the prototypical example (\ref{fform})
is not required for the general analysis: it is sufficient to
impose duality-covariant constraints, which in general may have different solutions.

Let us now turn to the constraints.
First, we require that $\eta^{MN}$ is an invariant tensor under the adjoint action with
$f^{M}{}_{NK}$,
 \be\label{invconstr}
  f^{(M}{}_{PK}\,\eta^{N)K} \ = \ 0\;.
 \ee
This is satisfied for (\ref{fform}) with $\eta^{MN}$ defined by (\ref{neweta}),
and we recall that the component $\eta_{\alpha\beta}$ is identified with the invariant Cartan-Killing form
of $G$. Together with the antisymmetry
of $f^{M}{}_{NK}$ in its lower indices, the constraint (\ref{invconstr}) implies that $f$ with
all indices raised or lowered with $\eta$ is totally antisymmetric,
 \be\label{fantisymm}
  f_{MNK} \ = \ f_{[MNK]}\;, \qquad f^{MNK} \ = \ f^{[MNK]}\;.
 \ee
Next, we require that
$f^{M}{}_{NK}$ satisfies the Jacobi identity
 \be\label{Jacobi}
  f^{M}{}_{N[K}\,f^{N}{}_{LP]} \ = \ 0\;,
 \ee
which is satisfied for (\ref{fform}) by virtue of the Jacobi identity for $f^{\alpha}{}_{\beta\gamma}$.

Apart from these algebraic constraints, we have to impose one new condition in addition to
the strong constraint (\ref{ODDconstr}): we require the differential constraint
 \be\label{fdelconstr}
  f^{M}{}_{NK}\,\partial_{M} \ = \ 0 \; ,
 \ee
when acting on fields or parameters.  By (\ref{fantisymm}) this implies that all derivatives
act trivially that are contracted with any index of $f^{M}{}_{NK}$. For the choice (\ref{fform})
this implies $\partial_{\alpha}=0$, as we will prove below.

To summarize, we impose the $O(D,D+n)$ covariant constraints (\ref{invconstr}), (\ref{Jacobi})
and (\ref{fdelconstr}). Any $f^{M}{}_{NK}$ satisfying these conditions will lead to a consistent, that is, gauge
invariant deformation of the abelian theory discussed above. A particular solution of these
constraints is given by (\ref{fform}) with $\partial_{\alpha}=0$ where, as we shall see below,
the theory reduces to the non-abelian low-energy action of the heterotic string. We stress,
however, that any solution obtained from this one by an $O(D,D+n)$ transformation also
satisfies the constraints. We will return to this point in sec.~4.

We close this section by introducing the modified or deformed gauge transformations.
Each $O(D,D+n)$ index will give rise to an adjoint rotation with the structure constants
$f^{M}{}_{NK}$.  In (\ref{newgauge}) we displayed this transformation for a tensor with an
upper index,
 \be\label{upperdef}
  \delta_{\xi}V^{M} \ = \ \widehat{\cal L}_{\xi}V^{M}-\xi^{N}f^{M}{}_{NK}V^{K}\;,
 \ee
and the transformation for a tensor with a lower index is given by
 \be
  \delta_{\xi}V_{M} \ = \ \widehat{\cal L}_{\xi}V_{M}+\xi^{K}f^{N}{}_{KM}V_{N}\;.
 \ee
This extends in a straightforward way to tensors with an arbitrary number of upper and lower
indices,  such that the generalized metric transforms as
 \be\label{deformgauge}
  \delta_{\xi}{\cal H}^{MN} \ = \  \widehat{\cal L}_{\xi}{\cal H}^{MN}
  -2\,\xi^{P}f^{(M}{}_{PK}\,{\cal H}^{N)K}\;.
\ee
By virtue of the constraints (\ref{invconstr}), the $O(D,D+n)$ invariant metric $\eta$ is
invariant under these transformations, $\delta_{\xi}\eta^{MN}=0$, which is a
generalization of the analogous property in the abelian case.
Moreover, the constraint (\ref{fdelconstr}) has two immediate consequences for these
deformed gauge transformations. First, the partial derivative of a scalar transforms covariantly,
 \be\label{covvarS}
  \delta_{\xi}(\partial_{M}S) \ = \ \widehat{\cal L}_{\xi}(\partial_{M}S) \ = \
  \widehat{\cal L}_{\xi}(\partial_{M}S)+\xi^{L}f^{K}{}_{LM}\partial_{K}S\;.
 \ee
Second,  any gauge transformation
with a parameter that is a gradient acts trivially,
 \be\label{trivial}
  \xi^{M} \ = \ \partial^{M}\chi \qquad\Rightarrow \qquad \delta_{\xi}{\cal H}^{MN} \ = \ 0\;,
 \ee
i.e., as for the abelian case there is a `gauge symmetry for gauge symmetries'.

\subsection{The non-abelian gauge transformations}
Let us now verify that the deformed gauge transformations (\ref{deformgauge}) indeed lead
to the required non-abelian gauge transformations if we choose (\ref{fform}) and set
$\tilde{\partial}^i = \partial_{\alpha} = 0$. The Yang-Mills gauge field transforms
as\footnote{In order to simplify the notation, we
assume from now on that the gauge coupling constant $g_0$ has been absorbed into the structure constants
$f^{\alpha}{}_{\beta\gamma}$, such that it does not appear explicitly in the formulas below.}
 \be\label{YangMills}
  \delta_{\Lambda}A_{i}{}^{\alpha} \ = \  \partial_{i}\Lambda^{\alpha}  + f^{\alpha}{}_{\beta\gamma}
  A_{i}{}^{\beta}\Lambda^{\gamma}\;.
 \ee
The $b$-field transforms according to (\ref{newdelb}) and thus its transformation rule is
not modified as compared to the abelian case.

We apply (\ref{deformgauge}) to particular components of ${\cal H}^{MN}$, where we
focus on the new terms proportional to $f^{M}{}_{NK}$, which we denote by $\delta^{\prime}$.
The variation of ${\cal H}^{ij}$ does not receive any modification since by (\ref{fform})
the $f$-dependent term in (\ref{deformgauge}) is zero for external indices $i,j$.
Thus, the metric $g_{ij}$ is still inert under $\Lambda$ transformations, as expected.
For components with external index $\alpha$, however, we find, e.g.,
 \be
  \delta^{\prime}_{\xi}{\cal H}^{i\alpha} \ = \ -g^{ik}\delta A_{k}{}^{\alpha} \ = \
  -\Lambda^{\beta}f^{\alpha}{}_{\beta\gamma}{\cal H}^{i\gamma} \quad \Rightarrow\quad
  \delta_{\Lambda}^{\prime}A_{k}{}^{\alpha} \ = \ f^{\alpha}{}_{\beta\gamma}A_{k}{}^{\beta}\Lambda^{\gamma}\;,
 \ee
which amounts to the required transformation rule (\ref{YangMills}).
Next, from ${\cal H}^{i}{}_{j}=-g^{ik}c_{kj}$ we infer that $\delta c_{ij}$
does not get corrected. In (\ref{c}) the symmetric combination quadratic in $A$
is invariant under the non-abelian part of (\ref{YangMills}), as one may easily confirm,
and therefore we conclude that also $\delta b_{ij}$ does not get modified as compared
to the abelian case, in agreement with (\ref{newdelb}). Thus, (\ref{deformgauge})
yields precisely the required gauge transformations.

In the remainder of this subsection, we discuss the closure of the deformed gauge transformations.
It is sufficient (and simplifies the analysis) to compute the closure on a vector $V^{M}$
whose gauge variation is given in (\ref{upperdef}).
The commutator of two such gauge transformations is then given by
 \begin{eqnarray}\nonumber
  \big[ \delta_{\xi_1},\delta_{\xi_2}\big] V^{M} &=& \delta_{\xi_1}\big(
  \xi_2^{N}\partial_{N}V^{M}+(\partial^{M}\xi_{2 N}-\partial_{N}\xi_2^{M})V^{N}
  -\xi_2^{K}f^{M}{}_{KN}V^{N}\big) -(1\leftrightarrow 2) \\ 
   &=&  \big[\widehat{\cal L}_{\xi_1},\widehat{\cal L}_{\xi_2}\big]V^{M}\\ \nonumber
   &&-\xi_2^{N}\partial_{N}
   \big(\xi_1^{K}f^{M}{}_{KP}V^{P}\big)
   -\big(\partial^{M}\xi_{2N}-\partial_{N}\xi_2^{M}\big)\xi_1^{K}
   f^{N}{}_{KP}V^{P}\\ \nonumber
   &&-\xi_2^{K}f^{M}{}_{KN}\big(\xi_1^{P}\partial_P V^{N}+(\partial^{N}\xi_{1P}-\partial_{P}\xi_1^{N})V^{P}
   -\xi_1^{P}f^{N}{}_{PQ}V^{Q}\big)
   -(1\leftrightarrow 2)\;.
 \end{eqnarray}
Using the constraints (\ref{fdelconstr}) and (\ref{Jacobi}) it is now relatively straightforward
to check that this can be rewritten as
 \be
   \big[ \delta_{\xi_1},\delta_{\xi_2}\big] V^{M} \ = \ \widehat{\cal L}_{\xi_{12}}V^{M}
   -\xi_{12}^{N} f^{M}{}_{NK}V^{K}\;,
 \ee
where
 \be\label{defCbracket}
   \xi_{12}^{M} \ = \ \xi_2^{N}\partial_{N}\xi_1^{M}-\frac{1}{2}\xi_{2N}\partial^{M}\xi_1^{N}
   - (1\leftrightarrow 2)-f^{M}{}_{NK}\xi_2^{N}\xi_1^{K}\;.
 \ee
Thus, we have verified the closure of the gauge algebra and thereby arrived at a generalization of the C-bracket that is deformed by the structure constants $f^{M}{}_{NK}$,
 \be\label{fbracket}
  \big[ X,Y\big]^{M}_{ f} \ = \ \big[X,Y\big]^{M}_{\rm C}-f^{M}{}_{NK}X^{N}Y^{K}\;.
 \ee

The C-bracket does not satisfy the Jacobi identities, but the resulting non-trivial Jacobiator
gives rise to a trivial gauge transformation that leaves the fields invariant.
The deformed bracket (\ref{fbracket}) has a similar property, which we investigate now.
First, we evaluate the Jacobiator,
 \be
  J_{f}(X,Y,Z) \ = \ \big[\big[ X,Y\big]_{f},Z\big]_{f}
  +\big[\big[ Y,Z\big]_{f},X\big]_{f}+\big[\big[ Z,X\big]_{f},Y\big]_{f}\;.
 \ee
We  compute from (\ref{fbracket})
 \be
  \begin{split}
    \big[\big[ X,Y\big]_{f},Z\big]_{f}^{M} \ = \ &\big[\big[ X,Y\big]_{\rm C},Z\big]_{\rm C}^{M}
    +f^{M}{}_{NK}f^{N}{}_{PQ}X^{P}Y^{Q}Z^{K} \\
    &+f^{M}{}_{NK}\big(Z^{P}\partial_{P}(X^{N}Y^{K})-(X^{P}\partial_{P}Y^{N}
    -Y^{P}\partial_{P}X^{N})Z^{K}\big) \\
    &+\frac{1}{2}f^{N}{}_{KL}\big(X^{K}Y^{L}\partial^{M}Z_{N}-Z_{N}\partial^{M}(X^{K}Y^{L})\big)\;,
  \end{split}
 \ee
where we used the constraint   (\ref{fdelconstr}).  Using the Jacobi identity (\ref{Jacobi})
we obtain after a brief computation
 \be\label{Jf}
  J_{f}(X,Y,Z)^{M} \ = \ J_{\rm C}(X,Y,Z)^{M} -\frac{1}{2}\partial^{M}\big(f_{NKL}X^{N}Y^{K}Z^{L}\big)\;.
 \ee
Here, $J_{\rm C}$ is the Jacobiator of the C-bracket, which has been proved in  \cite{Hull:2009zb}
to be a gradient. Thus, we infer from (\ref{Jf})
 \be
  J_{f}(X,Y,Z)^{M} \ = \ \partial^{M}\Big(\chi_{\rm C}(X,Y,Z)-\frac{1}{2}
  f_{NKL}X^{N}Y^{K}Z^{L}\Big)\;,
 \ee
where $\chi_{\rm C}$ is given in eq.~(8.29) of    \cite{Hull:2009zb}. We have seen in (\ref{trivial})
that a gauge parameter that takes the form of a pure gradient gives rise to a trivial gauge transformation
on the fields. Thus, in precise analogy to \cite{Hull:2009zb},
the non-vanishing Jacobiator is consistent with the fact that the
infinitesimal gauge transformations $\delta_{\xi}$ automatically satisfy the Jacobi identity.

We finally note that, in analogy to the discussion at the end of sec.~2.2,
the modified form of the gauge algebra is consistent with the closure property
 \be\label{naclosure}
  \big[ \delta_{\Lambda_1},\delta_{\Lambda_2}\big]b_{ij} \ = \ \big(\delta_{\tilde{\xi}}+\delta^{}_{\Lambda}\big)b_{ij}\;,
  \qquad
  \Lambda^{\alpha} \ = \  f^{\alpha}{}_{\beta\gamma}\Lambda_1^{\beta}\Lambda_2^{\gamma}\;,
 \ee
where $\tilde{\xi}_{i}$ is given by (\ref{norbracket}).
In the mathematical terminology of sec.~2.2, the closure property (\ref{defCbracket}) or (\ref{naclosure})
amounts to a further generalization of the Courant bracket, involving the structure of a non-abelian Lie algebra,
in that the term $[\Lambda_1,\Lambda_2]$ has to be added in the last line of (\ref{newbracket}).

\subsection{The non-abelian action}
Next, we construct a deformation of the double field theory action parametrized by the $f^{M}{}_{NK}$
in such a way that it is gauge invariant under (\ref{deformgauge}) and leads to the required
low-energy action. For this we will start from the action written in Einstein-Hilbert like form \cite{Hohm:2010pp},
  \be\label{actR}
   S \ = \  \int dx \, d\tilde x  \, e^{-2d} \, {\cal R}({\cal H},d)\;,
 \ee
where ${\cal R}({\cal H},d)$ is given by
 \be
 \label{simplerformR}
 \begin{split}
  {\cal R} \  \equiv \ &\;4\,{\cal H}^{MN}\partial_{M}\partial_{N}d
  -\partial_{M}\partial_{N}{\cal H}^{MN} \\[1.2ex]
   & -4\,{\cal H}^{MN}\partial_{M}d\,\partial_{N}d
   + 4 \partial_M {\cal H}^{MN}  \,\partial_Nd\;\\[1.0ex]
    ~&+\frac{1}{8}\,{\cal H}^{MN}\partial_{M}{\cal H}^{KL}\,
  \partial_{N}{\cal H}_{KL}-\frac{1}{2}{\cal H}^{MN}\partial_{M}{\cal H}^{KL}\,
  \partial_{K}{\cal H}_{NL}\;.
 \end{split}
 \ee
It is defined such that it is a scalar under generalized Lie derivatives,
 \be\label{Rscalar}
  \delta_{\xi}{\cal R} \ = \ \xi^{P}\partial_{P}{\cal R}\;,
 \ee
which, together with the gauge
variation (\ref{manifestH}) of the dilaton, implies gauge invariance of the action.
Here we modify the form of ${\cal R}$ such that (\ref{Rscalar}) be preserved
under the deformed gauge transformations (\ref{deformgauge}).

The result for the deformed scalar curvature is given by
 \be\label{defR}
 \begin{split}
  {\cal R}_f \ = \ {\cal R}&-\frac{1}{2}f^{M}{}_{NK}\,{\cal H}^{NP}{\cal H}^{KQ}\partial_{P}{\cal H}_{QM}\\
  &-\frac{1}{12} f^{M}{}_{KP}f^{N}{}_{LQ}{\cal H}_{MN}{\cal H}^{KL}{\cal H}^{PQ}
  -\frac{1}{4}f^{M}{}_{NK}f^{N}{}_{ML}{\cal H}^{KL}-\frac{1}{6}f^{MNK}f_{MNK}\;,
 \end{split}
 \ee
and reduces for the abelian case $f=0$ to the previous expression.
Remarkably, the structure in the second line is precisely analogous to the scalar
potential appearing for Kaluza-Klein reduction on group manifolds \cite{Lu:2006ah}
and, for instance, in ${\cal N}=4$ gauged supergravity in $D=4$
\cite{Schon:2006kz}.\footnote{In fact, the scalar potential in ${\cal N}=4$ gauged supergravity
for so-called electric gaugings is, up to an overall prefactor, precisely given by
the second line of (\ref{defR}), see eq.~(2.2) in \cite{ReidEdwards:2008rd}.}
We next verify that this action evaluated for (\ref{fform}) and $\tilde{\partial}^i=\partial_{\alpha}=0$
gives rise to the required non-abelian form of the low-energy action of the heterotic string.

The non-abelian field strength with structure
constants $f^{\alpha}{}_{\beta\gamma}$ is given by
 \be\label{explF}
  F_{ij}{}^{\alpha} \ = \ \partial_iA_{j}{}^{\alpha}-\partial_jA_{i}{}^{\alpha}+f^{\alpha}{}_{\beta\gamma}
  A_i{}^{\beta}A_j{}^{\gamma}\;,
 \ee
while the field strength of the $b$-field is modified by the Chern-Simons 3-form and thus reads
explicitly
 \be\label{explH}
  \hat{H}_{ijk} \ = \ 3\left(
  \partial_{[i} b_{jk]} -
  \kappa_{\alpha\beta}A_{[i}{}^{\alpha}\Big(\partial_j A_{k]}{}^{\beta}
  +\tfrac{1}{3}f^{\beta}{}_{\gamma\delta} A_{j}{}^{\gamma}A_{k]}{}^{\delta}\Big)\right)\;.
 \ee
We recall that here we do not indicate the gauge coupling constant explicitly,
but rather absorb it into the structures constants.
Using (\ref{explF}) and (\ref{explH}), the $f$-dependent non-abelian couplings in the
low-energy Lagrangian in (\ref{hetaction}) are found to be
\begin{eqnarray}\label{newterms}
  {\cal L}_{f} &=& -f_{\alpha \beta \gamma}
  \, g^{ik}\,g^{jl}\,\partial_i A_{j}{}^{\alpha}
  A_k{}^{\beta}A_l{}^{\gamma}
  -\frac{1}{4}  f^{\alpha} {}_{\beta \gamma}f_{\alpha \delta \epsilon }\,g^{ik}\,g^{jl}\,
  A_i{}^{\beta}A_{j}{}^{\gamma}A_{k}{}^{\delta}A_{l}{}^{\epsilon} \\
  \nonumber &&  + \ \frac{1}{2} f_{\alpha \beta \gamma}  \, g^{ik}\, g^{jl}  \, g^{pq} \, \partial_i b_{jp}  \,  A_k{}^{\alpha} \, A_l{}^{\beta} \, A_q{}^{\gamma} - \frac{1}{2}  f_{\alpha \beta \gamma} \, g^{ik} \, g^{jl} \, g^{pq} \, A_{i \delta} \, \partial_j A_{p}{}^{\delta} \, A_{k}{}^{\alpha} \, A_{l}{}^{ \beta} \, A_{q}{}^{\gamma} \\
  \nonumber && - \frac{1}{12}   f_{\alpha \beta \gamma}  f_{\delta \epsilon \zeta}  \, g^{ik} \, g^{jl} \, g^{pq} \,
  A_i{}^{\alpha} A_j{}^{\beta}A_{p}{}^{\gamma} A_{k}{}^{\delta}  A_{l}{}^{\epsilon} A_{q}{}^{\zeta}  \; ,
\end{eqnarray}
where the first line originates from the Yang-Mills terms
and the second and third line from the non-abelian parts of the Chern-Simons 3-form.

To evaluate the new terms in (\ref{defR}), we define
 \be\label{Ris}
  {\cal R}_{f} \ = \ {\cal R}  -\frac{1}{2}{\cal R}_1-\frac{1}{12}{\cal R}_2-\frac{1}{4}{\cal R}_3-\frac{1}{6}f^{MNK}f_{MNK}\;,
 \ee
where the ${\cal R}_i$ are the respective terms in  (\ref{defR}) (in the order given there).
Setting $\tilde{\partial}^i = \partial_{\alpha} = 0$, the first term yields
\begin{eqnarray}
 \nonumber {\cal R}_1 &=& f^{M}{}_{NK}\,{\cal H}^{NP}{\cal H}^{KQ}\partial_{P}{\cal H}_{QM} \ = \ f^{\alpha}{}_{\beta \gamma} \, {\cal H}^{\beta i} \,
\big[ {\cal H}^{\gamma j} \partial_i {\cal H}_{j \alpha} + {\cal H}^{\gamma}{}_{ j} \partial_i {\cal H}^{j}{}_{\alpha}  +  {\cal H}^{\gamma}{}_{ \delta} \partial_i {\cal H}^{\delta}{}_{\alpha}  \big] \\
\nonumber &=& f^{\alpha}{}_{\beta \gamma} \, (-g^{ik} A_{k}{}^{\beta}) \Big[ (-g^{jl} A_{l}{}^{\gamma}) \, \partial_i  (c_{pj} \, g^{pq} \, A_{q \alpha} + A_{j \alpha}) + (c_{pj} \, g^{pq} \, A_{q }{}^{\gamma} + A_{j}{}^{\gamma})  \, \partial_i (-g^{jl} A_{l \alpha}) \\
 &&+ (\delta^{\gamma}{}_{\delta} + A_{j}{}^{\gamma} \, g^{jl} \, A_{l \delta}) \, \partial_i
(A_{p}{}^{\delta} \, g^{pq} \, A_{q \alpha}) \Big] \, .
\end{eqnarray}
Similar to the computation for the abelian case, one can simplify the above terms separately for those involving $(\partial g)$ and those not having derivatives of the metric. The result is
\begin{eqnarray}
{\cal R}_1 \big|_{(\partial g)^1} &=& f^{\alpha}{}_{\beta \gamma} \, (-g^{ik} A_{k}{}^{\beta}) (- \partial_i g^{pq}) \Big[ g^{jl} A_{l}{}^{\gamma} A_{q \alpha} (c_{pj} + c_{jp} - A_{j \delta} A_{p}{}^{\delta}) + A_{p}{}^{\gamma} A_{q \alpha} -  A_{p}{}^{\gamma} A_{q \alpha} \Big]
\nonumber \\ &=& 0  \,\; ,
\end{eqnarray}
where the last equality follows from the definition of $c_{ij}$ in (\ref{c}), and
\begin{eqnarray}
{\cal R}_1 \big|_{(\partial g)^0} &=& 2 f_{\alpha \beta \gamma}
  \, g^{ik}\,g^{jl}\,\partial_i A_{j}{}^{\alpha}  A_k{}^{\beta}A_l{}^{\gamma} - f_{\alpha \beta \gamma}  \, g^{ik}\, g^{jl}  \, g^{pq} \, \partial_i b_{jp}  \,  A_k^{\alpha} \, A_l^{\beta} \, A_q^{\gamma}  \nonumber \\
  &&+  f_{\alpha \beta \gamma} \, g^{ik} \, g^{jl} \, g^{pq} \, A_{i \delta} \, \partial_j A_{p}{}^{\delta} \, A_{k}{}^{\alpha} \, A_{l}{}^{ \beta} \, A_{q}{}^{\gamma} \; .
\end{eqnarray}
Thus ${\cal R}_1$ yields the first, third, and fourth terms in (\ref{newterms})
if we choose the coefficients as in (\ref{Ris}). The other terms in (\ref{newterms})  do not contain any derivatives and hence they should be obtained from ${\cal R}_2$ and ${\cal R}_3$.
The computation for ${\cal R}_2$ and ${\cal R}_3$ is rather direct:
\begin{eqnarray}
 {\cal R}_2 &=& f^{M}{}_{KP}f^{N}{}_{LQ}{\cal H}_{MN}{\cal H}^{KL}{\cal H}^{PQ}
 \\ \nonumber
 &=& f_{\alpha \beta \gamma} \, f_{\delta \epsilon \zeta}\, (\delta^{\alpha \delta} + A_{i}{}^{\alpha} \, g^{ik} \, A_{k}{}^{\delta}) \, (\delta^{\beta \epsilon} + A_{j}{}^{\beta} \, g^{jl} \, A_{l}{}^{\epsilon}) \, (\delta^{\gamma \zeta} + A_{p}{}^{\gamma} \, g^{pq} \, A_{q}{}^{\zeta})  \\
 \nonumber &=& f_{\alpha \beta \gamma} \, f^{\alpha \beta \gamma} + 3 f_{\alpha \beta \gamma} \, f^{\alpha \beta}{}_{\delta} \, g^{ik} \, A_{i}{}^{\gamma}  \, A_{k}{}^{\delta} +  3 f_{\alpha \beta \gamma} \,
 f^{\alpha}{}_{ \delta \epsilon} \, g^{ik} g^{jl} \, A_{i}{}^{\beta} \, A_{j}{}^{\gamma}  \, A_{k}{}^{\delta}
 \, A_{l}{}^{\epsilon} \\
 \nonumber && + f_{\alpha \beta \gamma}  f_{\delta \epsilon \zeta}  \, g^{ik} \, g^{jl} \, g^{pq} \, A_i{}^{\alpha} A_j{}^{\beta}A_{p}{}^{\gamma} A_{k}{}^{\delta}  A_{l}{}^{\epsilon} A_{q}{}^{\zeta} \; ,
\end{eqnarray}
and
\begin{eqnarray}
 {\cal R}_3 &=& f^{M}{}_{NK}f^{N}{}_{ML}{\cal H}^{KL}  \\
 \nonumber &=& f^{\alpha}{}_{\beta \gamma} \, f^{\beta}{}_{\alpha \delta} (\delta^{\gamma \delta} + A_{i}{}^{\gamma} \, g^{ik} \, A_{k}{}^{\delta}) \ = \ - f_{\alpha \beta \gamma} \, f^{\alpha \beta \gamma} - f_{\alpha \beta \gamma} \, f^{\alpha \beta}{}_{\delta} \, g^{ik} \, A_{i}{}^{\gamma}  \, A_{k}{}^{\delta} \, ,
 \end{eqnarray}
where we have repeatedly used the total antisymmetry of $f_{\alpha \beta \gamma}$.
The coefficient of ${\cal R}_2$ in (\ref{Ris}) has been chosen such that
it matches the coefficient of the terms $f^2 A^6$. Moreover, in order to eliminate the term
$f^2 A^2$, which is not present in Yang-Mills theory, the coefficient of ${\cal R}_3$ is fixed to
be $-\frac{1}{4}$. Finally, in order to cancel the constant terms
$ f_{\alpha \beta \gamma} \, f^{\alpha \beta \gamma}$ in ${\cal R}_2$ and ${\cal R}_3$,
the last term in (\ref{Ris}) is required.
In total, we have verified that (\ref{defR}) induces precisely the correct non-abelian terms.

\subsection{Proof of gauge invariance}
We turn now to the proof that the deformed action defined by (\ref{defR}) is invariant under the
deformed gauge transformations (\ref{deformgauge}).
The unmodified ${\cal R}$ transforms as a scalar under the unmodified gauge transformations.
We have to prove that its variation under the modified part of the gauge transformation,
which is proportional to $f$, cancels against the variation of the new terms involving $f$.

Since all $O(D,D+n)$ indices are
properly contracted it is sufficient to focus on the subset of variations that are non-covariant
and which we will denote by $\Delta_{\xi}$.
Specifically, in ${\cal R}$ the new non-covariant contributions originate from partial derivatives only.
For instance, for the following structure
the $f$-dependent terms in the gauge variation, denoted by $\delta_{\xi}'$, read
 \be\label{deltaprime}
  \delta_{\xi}' \big( \partial_{M}{\cal H}^{KL}\big) \ = \
  \xi^{P}f^{Q}{}_{PM}\partial_{Q}{\cal H}^{KL}
  -2\xi^{P}f^{(K}{}_{PQ}\,\partial_{M}{\cal H}^{L)Q}-2\partial_{M}\xi^{P}f^{(K}{}_{PQ}{\cal H}^{L)Q}\;,
 \ee
where the first term has been added by hand, which is allowed since it is zero
by the constraint (\ref{fdelconstr}). The first two terms represent the covariant contributions,
while the last term is non-covariant. We thus find
 \be\label{noncoder}
  \Delta_{\xi}\big( \partial_{M}{\cal H}^{KL}\big) \ = \
  -2\partial_{M}\xi^{P}f^{(K}{}_{PQ}{\cal H}^{L)Q}\;.
 \ee
Since we saw that $\eta^{MN}$ can be viewed as an invariant tensor under the
modified gauge transformations (\ref{deformgauge}), we can derive from this result,
by lowering indices with $\eta$, the following form
 \be\label{noncov23}
  \Delta_{\xi}\big( \partial_{M}{\cal H}_{KL}\big) \ = \
  2\partial_{M}\xi^{P}f^{Q}{}_{P(K}{\cal H}_{L)Q}\;.
 \ee
Moreover, from (\ref{covvarS}) we infer
 \be
   \Delta_{\xi}(\partial_{M}d) \ = \ 0 \;.
 \ee
Using this and (\ref{noncoder}), it is straightforward to see that all dilaton-dependent terms in
(\ref{simplerformR}) are separately invariant under the deformed part of the
gauge transformations.  For instance
 \be
  \Delta_{\xi}\big(4\partial_{M}{\cal H}^{MN}\partial_{N}d\big) \ = \
  -8\partial_{M}\xi^{P}\,f^{(M}{}_{PQ}\,{\cal H}^{N)Q}\,\partial_{N}d \ = \  0
 \ee
easily follows with (\ref{fdelconstr}). All other $d$-dependent terms can also be seen
to be gauge invariant by virtue of (\ref{fdelconstr}). Similarly, the term involving a second derivative
of ${\cal H}$ is gauge invariant,
 \be
  \delta^{\prime}_{\xi}\left(\partial_{M} \partial_{N}{\cal H}^{MN}\right) \ = \
  - 2\partial_{M}\big(\xi^{P}f^{(M}{}_{PQ}\,\partial_{N}{\cal H}^{N)Q}+
  \partial_{N}\xi^{P}f^{(M}{}_{PQ}{\cal H}^{N)Q}\big) \ = \ 0\;,
 \ee
where (\ref{deltaprime}) has been used.
Thus, we have to focus only on the terms in the last line of (\ref{simplerformR}), whose variation
with a little work can be brought to the form
 \be\label{f1var}
  \Delta_{\xi}{\cal R} \ = \ -\frac{1}{2}\partial_{N}\xi^{L}f^{M}{}_{LK}{\cal H}^{NP}{\cal H}^{QK}
  \partial_{P}{\cal H}_{MQ}
  -\partial^{M}\xi_{L} f^{L}{}_{NK}{\cal H}^{NP}{\cal H}^{KQ}\partial_{P}{\cal H}_{QM}\;.
 \ee
These terms have to be cancelled by the variations of the new terms in ${\cal R}_{f}$.

There are various contributions to the gauge transformations of the $f$-dependent terms in (\ref{defR}).
First, the partial derivative of ${\cal H}$ in the first line transforms non-covariantly already under
the unmodified part of the gauge transformations, but it can be easily checked, using eq.~(4.36)
from \cite{Hohm:2010pp}, that this
contribution is zero by (\ref{fdelconstr}). Next, we have to keep in mind that $f^{M}{}_{NP}$
is constant and thus does not transform with a generalized Lie derivative with respect to $\xi^{M}$.
The resulting non-covariant terms can be accounted for by assigning a fictitious non-covariant
variation to $f$ (with the opposite sign),
 \be\label{noncovf}
  \Delta_{\xi}f^{M}{}_{NK} \ = \ -\widehat{\cal L}_{\xi}f^{M}{}_{NK} \  = \
  -\partial^{M}\xi_{P}\,f^{P}{}_{NK}-\partial_{N}\xi^{P}\,f^{M}{}_{PK}-\partial_{K}\xi^{P}\,f^{M}{}_{NP}\;,
 \ee
where the constancy of $f$ and (\ref{fdelconstr}) has been used in the final step.
Using this, the variation of the $f$-dependent term in the first line of (\ref{defR}) can be seen to
precisely cancel (\ref{f1var}), which in turn fixes the coefficient of this term in ${\cal R}_f$ uniquely.

Next, using (\ref{noncov23}), the term in the first line of  (\ref{defR}) gives a variation proportional
to $f^2$,
 \be
 \begin{split}
  -\frac{1}{2}\Delta_{\xi}\big(f^{M}{}_{NK}{\cal H}^{NP}{\cal H}^{KQ}\partial_{P}{\cal H}_{QM}\big)
  \ = \ &-\frac{1}{2}f^{M}{}_{NK}f^{L}{}_{RQ}\partial_{P}\xi^{R}{\cal H}_{ML}{\cal H}^{NP}{\cal H}^{KQ}\\
  &-\frac{1}{2}f^{M}{}_{NK}f^{K}{}_{RM}\partial_{P}\xi^{R}{\cal H}^{NP}\;.
 \end{split}
 \ee
Thus, we get two contributions: one cubic in ${\cal H}$ and one linear in ${\cal H}$. The cubic
term is cancelled by the variation of the first term in the second line of (\ref{defR})
according to (\ref{noncovf}), which in turn fixes the coefficient of this term.
The term linear in ${\cal H}$ is cancelled by the variation (\ref{noncovf})
of the second term in the second line of (\ref{defR}), which finally fixes the coefficient of this term.
The last term in (\ref{defR}) is constant and thus trivially gauge invariant.
In total, we have proved that the modified scalar curvature ${\cal R}_f$ transforms as in (\ref{Rscalar}),
i.e., as a scalar, under the deformed gauge transformations  (\ref{deformgauge}),
and thus that the Einstein-Hilbert like action (\ref{actR}) is gauge invariant.

\section{The covariant constraints and their solutions}
In this section we discuss the $O(D,D+n)$ covariant differential constraints (\ref{ODDconstr}) and
(\ref{fdelconstr0}) and their solutions. Before that, we explain the relation of (\ref{ODDconstr})
to the level-matching condition in string theory.

\subsection{Relation to level-matching condition}
In the abelian case, for which (\ref{fdelconstr0}) trivializes,  the remaining
constraint (\ref{ODDconstr}) has a rather direct relation to
the level-matching condition of closed string theory.
In the original double field theory construction for the bosonic string,
the level-matching requires
for the massless sector \cite{Hull:2009mi}
 \be
  L_0-\bar{L}_0 \ = \ -p_i w^i \ = \ 0\;,
 \ee
where $p_i$ and $w^i$ are the momenta and winding modes on the torus, respectively.
Upon Fourier transformation, this implies  that in string field theory
all fields and parameters need to be annihilated by the differential operator
$\tilde{\partial}^{i}\partial_{i}$. Here, we require the stronger form that also all products of
fields and parameters are annihilated. Similarly, the extended form (\ref{strong1}) and (\ref{strong2})
of the constraint is the stronger version of the level-matching condition in
heterotic string theory, which will be discussed next.

We start by recalling the (bosonic part of) the world-sheet action for
heterotic string theory, which is given by \cite{Narain:1986am}
 \be
  S \ = \ \frac{1}{2\pi}\int d\tau d\sigma\Big[ G_{ij}\partial_a X^{i}\partial^a X^{j}+\varepsilon^{ab}
  B_{ij}\partial_aX^{i} \partial_{b}X^{j}+\partial_a X_{\alpha}\partial^{a}X^{\alpha}
  +\varepsilon^{ab}A_{i \alpha}\partial_a X^{i}\partial_{b}X^{\alpha}\Big]\;.
 \ee
Here, $X^i\sim X^i+2\pi k^i$, $k^i\in \mathbb{Z}$, denotes the periodic coordinates of the torus,
and we have not displayed the non-compact coordinates. The $X^{\alpha}$ are 16 internal
left-moving coordinates, i.e., satisfying the constraint $(\partial_{\tau}-\partial_{\sigma})X^{\alpha}=0$.
In this subsection, the indices $a,b$ label the
world-sheet coordinates $\tau,\sigma$, and $G$, $B$ and $A$ are the
backgrounds. We split the world-sheet scalars into left- and right-moving parts,
$X^{i}=X^{i}_{L}+X^{i}_{R}$, whose zero-modes are
 \be
 \begin{split}
   X^{i}_L(\tau+\sigma) \ &= \ \tfrac{1}{2}x_0^{i}+\tfrac{1}{2}p^{i}_{L}(\tau+\sigma)\;, \\
   X^{i}_R(\tau-\sigma) \ &= \ \tfrac{1}{2}x_0^{i}+\tfrac{1}{2}p^{i}_{R}(\tau-\sigma)\;, \\
   X^{\alpha}(\tau+\sigma) \ &= \ x_0^{\alpha}+p_L^{\alpha}(\tau+\sigma)\;.
  \end{split}
  \ee
Following the canonical quantization of \cite{Narain:1986am} (see also the discussion
around eqs.~(11.6.17) in \cite{Polchinski:1998rr}),
the left- and right-moving momenta can in turn be written as
 \be\label{pLR}
  \begin{split}
   p_{L\,i} \ &= \ \tfrac{1}{2} p_i + \left(G_{ij}-B_{ij}\right) w^{j}-\tfrac{1}{2}A_{i \alpha}
   \left(q^{\alpha}+\tfrac{1}{2}A_{j}{}^{\alpha}w^{j}\right)\;, \\
   p_{R\,i} \ &= \ \tfrac{1}{2} p_i - \left(G_{ij}+B_{ij}\right) w^{j}-\tfrac{1}{2}A_{i \alpha}
   \left(q^{\alpha}+\tfrac{1}{2}A_{j}{}^{\alpha}w^{j}\right)\;, \\
   p_L^{\alpha} \ &= \ q^{\alpha}+A_{i}{}^{\alpha}w^i\;,
  \end{split}
 \ee
where the momentum and winding quantum numbers $p_i$ and $w^{i}$, respectively,
are integers as a consequence
of the periodicity of the $X^i$, while the $q^{\alpha}$ take values in the root lattice
of $E_8\times E_8$ or $SO(32)$.

Let us now turn to the level-matching condition, where for definiteness we work in the Green-Schwarz formalism.
We truncate to the massless subsector of the heterotic string spectrum with 16 abelian gauge fields, i.e.,
taking values in the Cartan subalgebra. In other words, we restrict to the massless spectrum with
$N = 0$ and $\bar{N} = 1$ and thereby truncate out the 480 remaining gauge fields,
which appear for $N = 0$ and $\bar{N} = 0$, were $N$ and $\bar{N}$ are the number operators.
The level-matching condition for this subsector is given by
 \be\label{levelm}
  L_0-\bar{L}_0+a_L-a_R \ = \  L_0-\bar{L}_0 +1\ = \ (p_R^i)^2 - (p_L^i)^2- (p_L^{\alpha})^2 \ = \ 0\;,
 \ee
where the normal ordering constants are $a_L=1$ and $a_R=0$.
Inserting (\ref{pLR}) into (\ref{levelm}), we obtain
 \be
   2p_i w^i + q^{\alpha} q_{\alpha} \ = \ 0\;.
 \ee
If we interpret the $q_{\alpha}$, like $p_{i}$ and $w^{i}$, as the Fourier numbers corresponding to
a torus, this condition translates in coordinate space precisely into the differential constraint
(\ref{strong1}).
More precisely, the $q^{\alpha}$ are vectors in
the root lattice of $E_8\times E_8$ or $SO(32)$ rather than $T^{16}$, but these are topologically
equivalent, and so we conclude that, in precise analogy to the case of bosonic string theory originally analyzed in \cite{Hull:2009mi},
the level-matching condition amounts to the differential constraint (\ref{strong1})
(and, correspondingly, (\ref{strong2}) represents the stronger form of this constraint).
We stress that the non-abelian case to be discussed in the next subsection
is conceptually very different because it requires formally
the introduction of 496 extra coordinates together with the novel constraint (\ref{fdelconstr0}), which
have no direct interpretation in the full string theory.

\subsection{Solutions of the constraints}

Next, we turn to the discussion of the solutions of the strong constraint. As in the bosonic string,
we will show that all solutions of this constraint are locally related via an $O(D,D+n)$ rotation to solutions
for which fields and parameters depend only on the $x^{i}$. To see this, consider
the Fourier expansion of all fields and parameters, denoted generically by $A$, which take the form
 \be
  A(x,\tilde{x},y) \ = \ A\,e^{i (p_i x^i+w^i \tilde{x}_{i}+q_{\alpha}y^{\alpha})}\;,
 \ee
where we indicated for simplicity only a single Fourier mode. The quantum numbers combine
into a vector of $O(D,D+n)$,
 \be
  P_{M} \ = \   \big(\,w^{i}\,,\; p_i\,, \; q_{\alpha}\,\big)\;.
 \ee
The strong constraint now implies that
 \be
  \eta^{MN}\,P_{M}^{\frak{a}}\, P_{N}^{\frak{b}} \ = \ 0 \; ,
 \ee
for all $\frak{a},\frak{b}$ (which label the Fourier modes of all fields and parameters). Thus, all momenta
are null and mutually orthogonal. In other words, they lie in a totally null or isotropic subspace
of $\mathbb{R}^{2D+n}$.  The canonical example of such a subspace is given by a
space with $w^i=q_{\alpha}=0$, corresponding to a situation where all fields
and parameters depend only on the $x^i$. Since the flat metric on $\mathbb{R}^{2D+n}$ has
signature $(D,D+n)$,
the maximal dimension of any isotropic subspace is $D$.
It is a rather general result, related to Witt's theorem (see the discussion and references in
\cite{Hohm:2010jy}), that all isotropic subspaces of the same dimension are related by
isometries of the full space, i.e., here they are related by $O(D,D+n)$
transformations. In particular, one can always find an $O(D,D+n)$ transformation to a T-duality frame where
$w^i=q_{\alpha}=0$ and therefore one can always rotate into a frame where fields
and parameters depend only on $x^i$, as we wanted to show.

Next, we discuss the general non-abelian theory. In this case, the global $O(D,D+n)$ symmetry is broken
by a choice of non-vanishing structure constants $f^{M}{}_{NK}$ and, therefore, we have no longer
all T-duality transformations to our disposal in order to rotate into a frame in which the fields
depend only on $x^i$. This is, however, compensated by the additional constraint (\ref{fdelconstr})
which eliminates further coordinates for non-vanishing structure constants.

To illustrate this point, suppose that we choose $f^{M}{}_{NK}$ as in (\ref{fform}), i.e., the
only non-vanishing components $f^{\alpha}{}_{\beta\gamma}$ are given by the structure
constants of a semi-simple Lie group $G$. We can view $G$ as
the subgroup of $SO(n)$ that leaves the tensor
$f^{\alpha}{}_{\beta\gamma}$ invariant,\footnote{Any compact $n$-dimensional Lie group $G$ can be
canonically embedded into $SO(n)$. If we denote the generators of $\frak{so}(n)$ by
$K^{\alpha\beta}=-K^{\beta\alpha}$, the generators $t^{\alpha}$ of $G$
are embedded as $t^{\alpha}=\tfrac{1}{2}f^{\alpha}{}_{\beta\gamma}K^{\beta\gamma}$.} and so the global symmetry group is then broken
to $O(D,D)\times G$, where we view $G$ as the global subgroup of the
gauge group.
The constraint (\ref{fdelconstr}) can now be multiplied with the
structure constants, which implies
 \be
  0 \ = \ f^{\gamma}{}_{\delta\alpha}\,f^{\delta}{}_{\gamma\beta}\,\partial^{\beta} \ = \
  -2\,\kappa_{\alpha\beta}\,\partial^{\beta}\;,
 \ee
where $\kappa_{\alpha\beta}$ is the Cartan-Killing form.   As $\kappa_{\alpha\beta}$
is invertible for a semi-simple Lie algebra, we conclude $\partial_{\alpha}=0$, i.e.,
the constraint implies that all fields are independent of $y^{\alpha}$. The unbroken
$O(D,D)$ transformations can then be used as above in order to rotate into a T-duality frame in
which the fields are independent of $\tilde{x}$. In total, the constraints are still sufficient
in order to guarantee that the dependence on the `unphysical' coordinates $\tilde{x}$ and $y$
is either eliminated directly or removable by a surviving T-duality transformation.

Let us now turn to a more general situation where $f^{M}{}_{NK}$ is of the form (\ref{fform}),
but with the gauge group $G$ having some $U(1)$ factors.   Suppose, the gauge group
is of the form
 \be
  G \ = \ U(1)^p\times G_0\;,
 \ee
where $G_0$ is semi-simple and embedded into $O(n-p)$. If we split the indices
accordingly, $\alpha=(\underline{\alpha},\bar{\alpha})$, with $\underline{\alpha}=1,\ldots, p$
and $\bar{\alpha}=1,\ldots, n-p$, the non-vanishing components of $f^{M}{}_{NK}$
are given by the structure constants $f^{\bar{\alpha}}{}_{\bar{\beta}\bar{\gamma}}$
of $G_0$. The constraint (\ref{fdelconstr}) implies in this case only $\partial_{\bar{\alpha}}=0$,
i.e., that the fields are independent of the $n-p$ coordinates $y^{\bar{\alpha}}$. The unbroken T-duality
group is, however, given by $O(D,D+p)$ and thus larger than in the previous example.
Therefore,  as in the above discussion of the abelian case, these transformations can be
used in order to rotate into a T-duality frame in which the fields are both independent of
$\tilde{x}_{i}$ but also of the remaining $p$ coordinates $y^{\underline{\alpha}}$.
Thus, the constraints and residual T-duality transformations are again sufficient in
order to remove the dependence on $\tilde{x}$ and $y$.

We finally note that by virtue of the $O(D,D+n)$ covariance of the constraints any $f^{M}{}_{NK}$
obtained from (\ref{fform}) by a duality transformation also solves the constraints.
Presumably, these have to be regarded as physically equivalent to (\ref{fform}) and thereby
to the conventional low-energy action of heterotic string theory. It remains to be
investigated, however, whether there are different solutions to the constraints.
This is particularly interesting in the context of (generalized) Kaluza-Klein
compactifications, where the fields are independent of some of the $x^{i}$ and
for which the differential constraints may allow for more general solutions.  We leave this to future work.

\section{Frame formulation}
Here, we reformulate the above results in a frame-like language in order to make contact
with the formalism developed by Siegel \cite{Siegel:1993th}, as has been done in \cite{Hohm:2010xe} for the
double field theory extension of the bosonic string. We first discuss the abelian case,
which is straightforward, and then turn to the non-abelian case which requires an extension
of the formalism. The non-abelian case was already  
mentioned by Siegel in \cite{Siegel:1993bj}. Specifically, 
this reference discusses a modification of the coefficients of anholonomy and  
a corresponding deformation of the C-bracket, and these results coincide with 
our results given in eqs.~(\ref{newomega}) and (\ref{fcovrel}) below.

\subsection{Frame fields and coset formulation}
The basic field in the formalism of Siegel is a vielbein or frame field $e_{A}{}^{M}$
that is a vector under gauge transformations parameterized by $\xi^{M}$ and which is subject
to local tangent space transformations indicated by the flat index $A$. In the present case,
the tangent space group is $GL(D)\times GL(D+n)$ and the index splits as $A=(a,\bar{a})$.
Using the frame field and $\eta_{MN}$, one can define a tangent-space metric of signature
$(D,D+n)$,
 \be\label{Hdefcov}
  {\cal G}_{AB} \ = \ e_{A}{}^{M}\,e_{B}{}^{N}\,\eta_{MN}\;,
 \ee
and the frame field is constrained to satisfy
 \be\label{offconstr}
  {\cal G}_{a\bar{b}} \ = \  0\;.
 \ee
Starting from this frame field and the local tangent space symmetry, one may
introduce connections for this gauge symmetry, impose covariant constraints
and construct invariant generalizations of the Ricci tensor and scalar curvature.
Rather than repeating this construction here, we will just mention in the following the
new aspects in the case of the heterotic string theory and refer to \cite{Siegel:1993th} and \cite{Hohm:2010xe}
for more details.

The generalized metric can be defined as follows
 \be\label{Hroot}
  {\cal H}^{MN} \ = \ 2{\cal G}^{\bar{a}\bar{b}}\,e_{\bar{a}}{}^{M}e_{\bar{b}}{}^{N}-\eta^{MN} \ = \
  -2{\cal G}^{ab}\,e_{a}{}^{M}e_{b}{}^{N}+\eta^{MN}\;,
 \ee
where the equivalence of the two definitions is a consequence of the constraint (\ref{offconstr}).
Next, it is convenient to gauge-fix the tangent space symmetry by setting ${\cal G}_{AB}$
equal to $\eta_{MN}$ (up to a similarity transformation, c.f.~the discussion after eq.~(5.22)
in \cite{Hohm:2010pp}), such that (\ref{Hdefcov}) and (\ref{Hroot}) imply \cite{Hohm:2010pp}
 \be\label{Hcoset}
    {\cal H}^{MN} \ = \ \delta^{AB}\,e_{A}{}^{M}\,e_{B}{}^{N}\;.
 \ee
This leaves a local $O(D)\times O(D+n)$ symmetry unbroken, and in this gauge we can think
of the frame field $e_{A}{}^{M}$ as a $O(D,D+n)$-valued coset representative that is
subject to local $O(D)\times O(D+n)$ transformations. Thus, this formulation can be
viewed as a generalized coset space construction based on $O(D,D+n)/(O(D)\times O(D+n))$,
in analogy to the structure appearing in dimensional reduction of heterotic supergravity \cite{Maharana:1992my}.
Fixing the local symmetry further, one may give explicit parametrizations of the frame field
$e_{A}{}^{M}$ in terms of the physical fields that give rise to the form
(\ref{Hgenup}) of ${\cal H}^{MN}$ according to (\ref{Hcoset}), see, e.g.,~eq.~(4.12) in \cite{Maharana:1992my}.

We turn now to the definition of the scalar curvature ${\cal R}$ that can be used to define an
invariant action as in (\ref{actR}). It can be written in terms of `generalized coefficients
of anholonomy' $\Omega_{AB}{}^{C}$ that are defined via the C-bracket (\ref{cbracketdef}),
 \be\label{Canholonomy}
  \big[ e_{A},e_{B}\big]_{\rm C}^{M} \ = \ \Omega_{AB}{}^{C}\,e_{C}{}^{M}\;.
 \ee
Defining\footnote{We note that we changed notation as compared to
\cite{Siegel:1993th,Hohm:2010xe}, where this quantity has been denoted by $f$, in order to
distinguish it from the structure constants.}
 \begin{equation}\label{Defh}
  h_{ABC} \ = \ (e_{A} e_{B}{}^{M}) e_{CM} \; ,
 \end{equation}
where $e_{A}=e_{A}{}^{M}\partial_{M}$, one obtains explicitly
 \begin{equation}
  \Omega_{ABC} \ = \ 2 h_{[AB] C} + h_{C[AB]} \ = \ h_{AB C} + h_{B C A} + h_{CAB} \ = \ 3 h_{[ABC]} \, .
 \end{equation}
Here we used that the gauge condition implies that ${\cal G}_{AB}$ is constant and therefore
$h_{ABC} = - h_{ACB}$ from the definition (\ref{Defh}). Finally, defining
 \be
  \tilde{\Omega}_A \ = \ \partial_{M}e_{A}{}^{M}-2e_A d\;,
 \ee
the scalar curvature is given by
\begin{equation} \label{ricci_scalar}
   {\cal R} \ = \ e_a \tilde{\Omega}^a + \frac{1}{2} \tilde{\Omega}_a{}^2  + \frac{1}{2} e_a e_b {\cal G}^{ab}   -
   \frac{1}{4} \Omega_{ a b \bar{c}} {}^2  -  \frac{1}{12} \Omega_{[abc]}{}^2
   + \frac{1}{8} e^a {\cal G}^{bc}\, e_b
   {\cal G}_{ac}\; .
 \end{equation}
In \cite{Hohm:2010xe} it has been verified that starting from this expression for ${\cal R}$ and using
the definition of ${\cal H}^{MN}$ in terms of the frame fields, this reduces precisely to the
form given above in (\ref{simplerformR}), up to an overall factor of $4$. This proof immediately generalizes to the abelian
case of the heterotic string, as all expressions, including the definition (\ref{Hroot}) of ${\cal H}^{MN}$, are formally
the same.

\subsection{Non-abelian extension}
Let us now turn to the non-abelian generalization, which has also been 
mentioned in \cite{Siegel:1993bj}. A natural starting point is the deformed bracket
(\ref{fbracket}) of gauge transformations. We further generalize the
coefficients of anholonomy by defining
 \be\label{widehat}
  \big[ e_{A},e_{B}\big]_{f}^{M} \ = \ \widehat{\Omega}_{AB}{}^{C}e_{C}{}^{M}\;.
 \ee
By (\ref{fbracket}) and (\ref{Canholonomy}) this implies
 \be\label{newomega}
  \widehat{\Omega}_{AB}{}^{C} \ = \ \Omega_{AB}{}^{C}-f^{C}{}_{AB}\;,\qquad
  f^{C}{}_{AB} \ = \ f^{M}{}_{NK}\,e_{M}{}^{C}\,e_{A}{}^{N}\,e_{B}{}^{K}\;,
 \ee
where we introduced structure constants with flattened indices. The $f$-bracket
of two vectors that transform covariantly under the deformed gauge transformations
transforms covariantly in the same sense, i.e.,
 \begin{eqnarray}\label{fcovrel}
  \delta_{\xi} \big[ X,Y\big]^{M}_{ f}
  &=& \widehat{\cal L}_{\xi} \big[ X,Y\big]^{M}_{ f} - \xi^N f^{M}{}_{NK} \big[ X,Y\big]^{K}_{ f} \; .
 \end{eqnarray}
To see this, we recall from \cite{Hohm:2010xe}
that the C-bracket is invariant under the generalized Lie derivative. Thus, it
remains to be shown that the non-covariant part of the variation of the C-bracket
due to the deformed gauge variation cancels against the variation of the new
term in the $f$-bracket. As in the proof of gauge invariance of the action above,
we denote the non-covariant part of the variation by $\Delta_{\xi}$ and compute
 \begin{eqnarray}
  \Delta_{\xi} \big[ X,Y\big]^{M}_{ \rm  C} &=& - \xi^P f^{N}{}_{PK} X^K \partial_N Y^M + \frac{1}{2} \xi^P f^{N}{}_{PK} X^K \partial^M Y_N  \\
  \nonumber &&  - \ X^N \partial_N \big( \xi^P f^{M}{}_{PK} Y^K \big) +  \frac{1}{2} X^N \partial^M \big( \xi^P f_{NP}{}^{K} Y_K \big) - \{ X \leftrightarrow Y \} \, .
 \end{eqnarray}
Using the constraint (\ref{fdelconstr}), it is straightforward to verify that this
can be rewritten as
 \begin{equation}
  \Delta_{\xi} \big[ X,Y\big]^{M}_{ \rm  C} \ = \
  - \xi^N f^{M}{}_{NK} \big[ X,Y\big]^{K}_{ \rm  C} - (\widehat{\cal L}_{\xi} f^{M}{}_{NK}) X^N Y^K \; .
 \end{equation}
The second term here is precisely cancelled by the non-covariant variation of the $f$-dependent
term in the $f$-bracket, which finally proves the covariance relation (\ref{fcovrel}).

Next, we discuss the extension of the scalar curvature (\ref{ricci_scalar}). Given
the covariance of the $f$-bracket, it follows from (\ref{widehat}) that $\widehat{\Omega}$
is a scalar under $\xi^{M}$ transformations, while its frame transformations are
as in the abelian case. Therefore, if we replace in (\ref{ricci_scalar}) $\Omega$ by
$\widehat{\Omega}$, the resulting expression will also be a scalar. In the following
we will show that
 \begin{equation}\label{ricci_scalarf}
   {\cal R}_{f} \ := \ e_a \tilde{\Omega}^a + \frac{1}{2} \tilde{\Omega}_a{}^2  + \frac{1}{2} e_a e_b {\cal G}^{ab}   -
   \frac{1}{4} \widehat{\Omega}_{ a b \bar{c}} {}^2  -  \frac{1}{12} \widehat{\Omega}_{[abc]}{}^2
   + \frac{1}{8} e^a {\cal G}^{bc}\, e_b
   {\cal G}_{ac}\;
 \end{equation}
indeed agrees with the definition (\ref{defR}) above.

Inserting here the definition (\ref{newomega}), we infer
 \begin{equation} \label{action_frame}
 {\cal R}_f  \ = \ {\cal R} - \frac{1}{4} \big( -2 \Omega_{a b \bar{c}} f^{a b \bar{c}}
 + f_{a b \bar{c}} f^{a b \bar{c}}  \big) - \frac{1}{12} \big( -2 \Omega_{[a b c]} f^{a b c} + f_{a b c} f^{a b c} \big) \; .
 \end{equation}
Next, we rewrite these new contributions in terms of the generalized metric,
using the definition (\ref{Hroot}), which we rewrite here as
 \begin{eqnarray}
   e_{a}{}^M e^{ a N} &=& \frac{1}{2} \big(\eta^{MN} - {\cal H}^{MN}\big)  \; , \\
   e_{\bar{a}}{}^M e^{ \bar{a} N} &=& \frac{1}{2} \big( \eta^{MN} +{\cal H}^{MN}\big) \; .
\end{eqnarray}
The second term in (\ref{action_frame}) can be written as
\begin{eqnarray}
\frac{1}{2} \Omega_{a b \bar{c}} f^{a b \bar{c}} &=& \frac{1}{2} \big( h_{ab\bar{c}} + h_{b\bar{c} a} + h_{\bar{c} ab} \big) f^{ab \bar{c}}
\ = \ \frac{1}{2}\big(2 h_{ab \bar{c}} + h_{\bar{c} ab}\big) f^{ab \bar{c}} \; ,
\end{eqnarray}
where
\begin{eqnarray}\nonumber
 h_{ab \bar{c}} f^{ab \bar{c}} &=& e_{a}{}^{N} \partial_N e_{b}{}^{M} \, e_{\bar{c} M} e^{a K} e^{b P} e^{\bar{c}}{}_{Q} f_{KP}{}^{Q} \ = \ - \partial_N \big(e_{\bar{c} M} e^{\bar{c}}{}_{ Q}  \big) e_{a}{}^{N} e^{a K}  e_{b}{}^{M}  e^{b P} f_{KP}{}^{Q} \\
   &=& - \frac{1}{8} \,  \partial_N {\cal H}_{MQ} \big(\eta^{NK} - {\cal H}^{NK}\big)
   \big(\eta^{MP} - {\cal H}^{MP}\big) f_{KP}{}^{Q} \\
 \nonumber &=& - \frac{1}{8} \ f_{KP}{}^{Q} \, {\cal H}^{NK} {\cal H}^{MP} \partial_N {\cal H}_{MQ} \; .
\end{eqnarray}
The fourth term in (\ref{action_frame}) is given by
\begin{eqnarray}
\frac{1}{6} \Omega_{[abc]} f^{abc} &=&  \frac{1}{2} h_{[abc]} f^{abc} \ = \ \frac{1}{2} h_{cab} f^{abc} \; ,
\end{eqnarray}
where in the last step the total antisymmetry of $f_{MNK}$ has been used.
Then, adding the second and fourth term, we obtain
\begin{eqnarray}
\frac{1}{2} \Omega_{a b \bar{c}} f^{a b \bar{c}} + \frac{1}{6} \Omega_{[abc]} f^{abc} &=&  h_{ab \bar{c}}  f^{ab \bar{c}} + \frac{1}{2} \big( h_{\bar{c} ab}  f^{ab \bar{c}} + h_{cab} f^{abc}  \big) \\
\nonumber &=&  h_{ab \bar{c}}  f^{ab \bar{c}} +   \frac{1}{2} h_{C a b} f^{a b C} \; ,
\end{eqnarray}
where
\begin{equation}
h_{ C a b} f^{ a b C} \ = \  e_{C}{}^{N} \partial_N e_{a}{}^{M} \, e_{b M} e^{a K} e^{b P} e^{C}{}_{Q} \, f_{KP}{}^{Q}  = f_{KP}{}^{Q}\, \partial_Q e_{a}{}^{M}\, e^{a K} e_{b M}  e^{b P} \ = \ 0 \; .
\end{equation}
The third and the fifth term in (\ref{action_frame}) can be evaluated directly. The third term yields
\begin{eqnarray}
-\frac{1}{4} f_{ab \bar{c}} f^{ab \bar{c}} &=&  -\frac{1}{4} e_{a}{}^{M} e_{b}{}^{K} e_{\bar{c}}{}^{P} e^{a N} e^{b L} e^{\bar{c} Q} \, f_{MKP} \, f_{NLQ} \\
\nonumber &=& - \frac{1}{32} \big(\eta^{MN} - {\cal H}^{MN}\big) \big(\eta^{KL} - {\cal H}^{KL}\big)
\big(\eta^{PQ} + {\cal H}^{PQ}\big) \, f_{MKP} \, f_{NLQ} \\
\nonumber &=& - \frac{1}{32} \Big[  f_{MNP} \,  f^{MNP} - {\cal H}^{MN} f_{MKP} \, f_{N}{}^{KP} - {\cal H}^{KL} {\cal H}^{PQ} \, f_{MKP} \, f^{M}{}_{LQ} \\
\nonumber && \qquad\;\; + {\cal H}^{MN} {\cal H}^{KL} {\cal H}^{PQ} \, f_{MKP} \, f_{NLQ}  \Big] \; ,
\end{eqnarray}
while the fifth term reads
\begin{eqnarray}
-\frac{1}{12} f_{ab c} f^{ab c} &=&  -\frac{1}{12} e_{a}{}^{M} e_{b}{}^{K} e_{c}{}^{P} e^{a N} e^{b L} e^{c Q} \, f_{MKP} \, f_{NLQ} \\
\nonumber &=& -  \frac{1}{96} \big(\eta^{MN} - {\cal H}^{MN}\big) \big(\eta^{KL} - {\cal H}^{KL}\big)
\big(\eta^{PQ} - {\cal H}^{PQ}\big) \, f_{MKP} \, f_{NLQ} \\
\nonumber &=& - \frac{1}{96}  \Big[  f_{MNP} \,  f^{MNP}  - 3 {\cal H}^{MN} f_{MKP} \, f_{N}{}^{KP}  + 3 {\cal H}^{KL} {\cal H}^{PQ} \, f_{MKP} \, f^{M}{}_{LQ}  \\
\nonumber && \qquad\;\; - {\cal H}^{MN} {\cal H}^{KL} {\cal H}^{PQ} \, f_{MKP} \, f_{NLQ}  \Big] \; .
\end{eqnarray}
Finally, combining all contributions, they agree precisely with the required form in terms
of ${\cal H}^{MN}$, up to the same overall factor of $4$ that arises in the abelian case, c.f.~\cite{Hohm:2010xe}.

\section{Conclusions and Outlook}
In this paper we have extended the double field theory formulation of \cite{Hohm:2010pp}
to the low-energy action
of the heterotic string, which features extra non-abelian gauge fields.  These
extra gauge fields neatly assemble with the massless
fields of closed bosonic string theory into an enlarged generalized metric that transforms
covariantly under the enhanced T-duality group $O(D,D+n)$ and thereby represent a
further `unification'. For the abelian subsector, the action takes the same structural form
as for the bosonic string, but based on the enlarged generalized metric.
In the non-abelian case, the T-duality group is broken to a subgroup, but interestingly the
action can still be written in a covariant fashion, with new couplings which are precisely analogous
to those encountered in lower-dimensional gauged supergravities. These new couplings are parametrized
by a tensor $f^{M}{}_{NK}$, and any such tensor satisfying a number of covariant constraints
defines a consistent deformation of the abelian theory.
This means that rather than having a proper global $O(D,D+n)$ symmetry, there is an
action of this group on the `space of consistent deformations' of the abelian theory.
Whether this space consists of a single $O(D,D+n)$ orbit or whether there are
more general solutions to the constraints that are inequivalent to (\ref{fform}) (and thereby
to the conventional Yang-Mills-type theory) remains
to be seen.

Several aspects of these results deserve further investigations. First, the gauge algebra
gives rise to a generalization of the Courant bracket when the dependence on the extra coordinates
is dropped such that the additional gauge structure enters non-trivially.
While extensions of the Courant bracket have been studied in the literature,
especially in the context of `exceptional generalized geometry' (see, e.g.,
\cite{Hull:2007zu,Baraglia:2011dg}), we are not aware of investigations of the structures
discussed here, and so it would be interesting to further study their mathematical aspects.
Moreover, general properties of gauged supergravities feature prominently in the
literature on `non-geometric compactifications' (see, e.g., \cite{ReidEdwards:2008rd})
as the most general gauged supergravities cannot be obtained
by any conventional Kaluza-Klein type reduction from higher-dimensional theories,
therefore requiring a sufficiently `non-geometric' novel framework.
As the construction presented here exhibits several features reminiscent to gauged
supergravity prior to any dimensional reduction, one might expect that
this theory can provide such a framework. We hope to return to these issues
in the near future.

\subsection*{Acknowledgments}
We are happy to acknowledge helpful discussions and correspondence with Marco Gualtieri,
Michael Haack,
Chris Hull, Ivo Sachs, Henning Samtleben, Michael Schulz  and especially Barton Zwiebach.

This work is supported by the U.S. Department of Energy (DoE) under the cooperative
research agreement DE-FG02-05ER41360. The work of OH is supported by the DFG -- The German Science Foundation. The work of SK is supported in part by a Samsung Scholarship.

\end{document}